\documentclass{ws-ijmpd}

\begin{document}

\markboth{L\'opez-Corredoira}
{Angular size test}

%
\catchline{}{}{}{}{}
%

\title{ANGULAR SIZE TEST ON THE EXPANSION OF THE UNIVERSE}

\author{MART\'IN L\'OPEZ-CORREDOIRA}

\address{Instituto de Astrof\'\i sica de Canarias,\\ 
E-38200 La Laguna, Tenerife, Spain\\
and\\
Departamento de Astrof\'\i sica, Universidad de La Laguna,\\
E-38205 La Laguna, Tenerife, Spain;\\
martinlc@iac.es}

\maketitle

\begin{history}
\received{Day Month Year}
\revised{Day Month Year}
\comby{Managing Editor}
\end{history}

\begin{abstract}
Assuming the standard cosmological model as correct,
the average linear size of galaxies with the same luminosity is
six times smaller at $z=3.2$ than at $z=0$, and their average angular size 
for a given luminosity is approximately proportional to $z^{-1}$.
Neither the hypothesis that galaxies which formed earlier have much
higher densities nor their luminosity evolution, mergers ratio,
or massive outflows due to a quasar feedback mechanism
are enough to justify such a strong size evolution. Also, at high redshift, 
the intrinsic ultraviolet surface brightness would be prohibitively 
high with this evolution, and the velocity dispersion much higher than
observed. We explore here 
another possibility to overcome this problem by
considering different cosmological scenarios that might make the
observed angular sizes compatible with a weaker evolution.

One of the models explored, a very simple phenomenological extrapolation of the
linear Hubble law in a Euclidean static universe,
fits the angular size vs. redshift dependence quite well, which is
also approximately proportional to $z^{-1}$ with this cosmological model.
There are no free parameters derived ad hoc, although the error bars
allow a slight size/luminosity evolution. 
The type Ia supernovae Hubble diagram can also be 
explained in terms of this model with no ad hoc fitted parameter.

WARNING: I do not argue here that the true Universe is static. My intention is just to
discuss which 
theoretical models provide a better fit to the data of 
observational cosmology.
\end{abstract}


\section{Introduction}

The analysis of the dependence of the angular size of some sources
with redshift was for many decades one of the most important 
geometric tests of cosmological models. Different cosmologies
predict different dependences for a given linear size and this
can be compared with the data from observations. The test, first
conceived by Hoyle\cite{Hoy59},
is simple in principle but its application is not so simple
because of the difficulty in finding a standard rod, a type of object
with no evolution in linear size over the lifetime of the Universe.

It is well known that the application of the
angular size ($\theta $) vs. redshift ($z$) test 
gives  a rough dependence of $\theta \propto z^{-1}$ for QSOs and
radio galaxies at radio wavelengths\cite{Mil71,Kel72,War74,Kap77,Kap87,Uba93}, 
for first ranked cluster galaxies in the
optical\cite{San72,Djo81,Pas96,Sch97}, and for the
separation of brightest galaxies in clusters \cite{LaV86}
or in QSO-galaxy pairs of the same redshift\cite{Sap99}.
The deficit of large objects at high redshifts with respect to
the predictions of an expanding Universe is believed to
be an evolutionary effect by which galaxies were smaller in the
past (e.g., \cite{Mil71}), or a selection effect (e.g., \cite{Jac73}). Thus, 
the $\theta \propto z^{-1}$ relationship, as predicted
by a static Euclidean Universe, would be just
a fortuitous coincidence of the superposition of the $\theta (z)$
dependence in the expanding Universe and evolutionary/selection effects.

Some other studies have tried to find better standard rods. 
Ultra-compact radio sources\cite{Kel93,Jac97,Gur99,Jac04,Jac06} 
were used to carry out an angular size test: a dependence 
different from $\theta \propto z^{-1}$ and 
closer to the predictions of expanding Universe models was found. The test
was even used to ascertain not only whether or not the Universe is expanding
but also to constrain the different cosmological parameters. However,
these applications are not free from selection effects\cite{Jac04} and,
as will be discussed in \S \ref{.ultracom}, interpretation of
the results of these tests is not so straightforward.

Another proposal\cite{Mar08a}  used the 
rotation speed of high redshift galaxies as a standard size indicator 
since there is a correlation between size and rotational velocity of galactic 
disks. This method was indeed applied\cite{Mar08b} over
a sample of emission-line galaxies with $0.2<z<1$, with the result that
Einstein--de Sitter cosmology is excluded to within 2-$\sigma $, and that
small galaxies (with fixed rotation velocity lower than 100 km/s) 
should show a strong evolution in size (at $z=1$ a factor two smaller than 
at $z=0$) and no evolution in luminosity within the concordance cosmology, 
while large galaxies do not evolve significantly, either in size or in 
luminosity. Unfortunately, this method requires spectroscopy, so the
sample has an upper limit of $z\sim 1$ even with very large telescopes,
and the uncertainties are so large that they cannot be interpreted directly 
without certain assumptions concerning evolution models. In principle, looking
at figures 4 and 5 of Marinoni {\it et al.}\cite{Mar08b}, one sees no reason to
exclude a $\theta \propto z^{-1}$ dependence.
 
The aim of this paper is to repeat the angular size test
for galaxies within a wide range of redshifts ($z=0.2-3.2$)
in optical--near infrared surveys (equivalent to the optical at rest).
Data from high spatial resolution surveys available nowadays, such as those
carried out with Hubble Telescope or FIRES, provide useful input for
this old test of the angular size with new analyses and interpretations.
Recent analysis of these data\cite{McI05,Bar05,Tru06} 
has shown that the linear size of the galaxies
with the same luminosity should be much lower than locally. 
However, this is true only if we considered the standard cosmological model 
as correct. In the present paper, I will do a reanalysis of these
data and consider different cosmological scenarios, 
which will shed further light on the degeneracy between 
expansion + evolution and non-expansion.

\section{Data}

Angular effective radii, defined as circularized 
S\'ersic half-light radii within which 50\% of the light is present, are
taken from  \cite{McI05,Bar05} for galaxies with $0.2<z<1$ and
from Trujillo {\it et al.}\cite{Tru06} for galaxies with $z>1$. 
Both samples separate approximately
early-type and late-type galaxies by means of the exponent ($n_S$) 
of their S\'ersic profile: $n_S>2.5$ for early-type galaxies and $n_S<2.5$ for
late type galaxies.

\cite{McI05,Bar05} provide data and angular size
measurement of the GEMS survey with two Hubble Space Telescope colors (F606W and
F850LP). In total, they have 929 galaxies.
The data processing and photometry are discussed by Rix {\it et al.}\cite{Rix04}.

Trujillo {\it et al.}\cite{Tru06} use near-infrared  
FIRES data of $z>1$ galaxies in the HDF-S and
MS 1054-03 fields to derive the angular size in the rest-frame V filter.
In total, there are 248 galaxies with $1<z<3.2$. There are 14 more galaxies
with $z>3.2$, but they are very few, very luminous, and their
photometric redshift determination was not very accurate; indeed,
the available on-line data through the FIRES Website, 
http://www.strw.leidenuniv.nl/~fires, gives new recalculated
redshifts and some of them are very different. 
The data processing and photometry are discussed 
in detail by Labb\'e {\it et al.}\cite{Lab03} for HDF-S and F\"orster-Schreiber
{\it et al.}\cite{For06} for the MS 1054-03 field. 

From these galaxies, I take only those with $3.4\times
10^{10}<L_{V,rest}(L_{\odot ,V})<2.5\times 10^{11}$, 
a total of 393 galaxies (271 galaxies from GEMS with $z<1$, and
122 galaxies from HDF-S/MS 1054-03 fields with $z>1$). 
The lower limit is the same as that adopted by
Trujillo {\it et al.}\cite{Tru06}, avoiding the
faintest ones in order to have a more homogeneous sample. 
The maximum limit is to avoid the galaxies away from the range
of local galaxies for which the relationship between radius and
luminosity was explored (explained in \S \ref{.astest}).
This is an almost
complete sample for redshifts up to $\approx 2.5$ \cite{Tru06}, 
and is incomplete for $2.5<z<3.2$. As we will see
later, it is unimportant whether the sample is complete or
not since the test is independent of the luminosity of the
galaxies, but within this restriction I will concentrate on 
the analysis of the brightest galaxies. 
A higher limit at the lowest luminosity (the sample would be complete up to
$z=3.2$ for $L_{V,rest}>6.7\times 10^{10}$ L$_\odot $)
would reduce  the number of galaxies too severely and the statistics 
would be poorer. In any case, I will also comment on the results
for these higher luminosity lower limits (see \S \ref{.results}).

\section{Angular size test}
\label{.astest}

I am going to analyze the variation of the angular size,
$\theta _{V,rest}$, of the galaxy rest-frame V with the redshift.
The angular size in the $z<1$ sample was measured at $\lambda _0=9450$
\AA \ (filter F850LP), $\theta_{\lambda _0}$, which,
due to the color gradients, is slightly different from the
angular size at V-rest. This difference is small\cite{McI05,Bar05} 
but I apply the following correction to it:

\begin{equation}
\theta_{V,rest}=\left \{ \begin{array}{ll}
        \theta_{\lambda _0}
	\left[1-0.11\left(\frac{\lambda _V(1+z)}{\lambda _0}-1\right)\right],
	& \mbox{ $n_S>2.5$} \\

        \theta_{\lambda _0}
	\left[1-0.08\left(\frac{\lambda _V(1+z)}{\lambda _0}-1\right)\right],
	& \mbox{ $n_S<2.5$}
\end{array}
\right \}
,\end{equation} 
with $\lambda_0=9450$ \AA , and $\lambda _V=5500$ \AA .
In any case, this correction is very small (less than 4\%)
and the results would not change significantly if it were not applied.
In HDF-S/MS 1054-03 data, the size was already measured
in the filter which gives V at rest\cite{Tru06}, so no correction is necessary.

Since we have a wide range of luminosities and types of galaxies, 
there is a huge dispersion of sizes for a given redshift, with
Malmquist bias, but this dispersion and bias can 
be reduced by defining $\theta _*$ as the equivalent angular size if
the galaxy were  early-type  with a given 
V-rest luminosity (I take $10^{10}$ L$_{\odot ,V}$; however, 
the variation of this number does not affect any result, just
the calibration in size):
\begin{equation}
\theta _*\equiv 
\theta \frac{R(10^{10}\ {\rm L_{\odot ,V}},n_S>2.5)}
{R(L_{V,rest},n_S)}
\label{thetaequiv}
,\end{equation}
where $R(L_{V,rest},n_S)$ is the average radius of
a galaxy for a given luminosity and exponent ($n_S$) 
of the S\'ersic profile. The number $n_S$ is affected by an important
uncertainty for galaxies with very small angular size ($\theta <0.125"$)
\cite{Tru06}, which produces some extra dispersion.
In any case, the dispersion of $R$ values  is moderate\cite{She03}, 
and, given that $\theta _*(z)$ does not contain the dispersion of
luminosities, it will present a much lower dispersion  than $\theta (z)$.
Certainly, $\theta _*(z)$ contains the spread of $\theta $ and $R$ so the
dispersion due to random errors in these quantities is larger for $\theta _*$ than
for $\theta $; but the dispersion due to the spread of luminosities dominates,
so, as said, $\theta _*$ will present a dispersion much lower than $\theta (z)$.
This definition also avoids selection effects due to Malmquist bias,
since $\theta _*$ for a given redshift should be nearly 
independent of the luminosity of  the galaxy, at least on average.

Shen {\it et al.}\cite{She03}(Eqs. 14-15) give the median
radius of a galaxy of given $r'_{SDSS}$-luminosity (K-correction applied) 
and $n_S$ for local SDSS galaxies.
In total, the radius (measured for $z\approx 0.1$ in the r-band,
which is more or less equivalent to V-band at rest) as a function of the 
absolute magnitude in $r'_{SDSS}$-rest $M_{r'-SDSS}$ is:

\begin{equation}
R(M_{r'-SDSS},n_S)
\label{shen}
\end{equation}\[
=\left \{ \begin{array}{ll}
        10^{-0.260M_{r'-SDSS}-5.06},& \mbox{ $n_S>2.5$} \\

        10^{-0.104M_{r'-SDSS}-1.71}[1+10^{-0.4(M_{r'-SDSS}+20.91)}]^{0.25},
	& \mbox{ $n_S<2.5$}
\end{array}
\right \}  \ {\rm kpc}
\]
We are not considering the evolution in this Eq. (\ref{shen});
all discussion of the effects of  evolution will be considered in
\S \ref{.evol}.
To translate this relationship into a
V-rest luminosity, we must make a color correction, as
in McIntosh {\it et al.}\cite{McI05} [however Trujillo {\it et al.}\cite{Tru06} 
 interpolate between the rest-frame g-band and
r-band]. Taking into account that
$\langle (V-r'_{SDSS,AB})\rangle$=0.33 for early-type galaxies
and $=0.30$ for late-type galaxies\cite{Fuk95} 
(this already includes the transformation of Vega to AB system;
no evolution is considered here),
and that $M_{V,\odot }=+4.79$ (Vega system) we will
have ($L_V$ in units of $10^{10}$ L$_{\odot, V}$)

\begin{equation}
R(L_V,n_S)=\left \{ \begin{array}{ll}
        A_0L_V^{0.65},& \mbox{ $n_S>2.5$} \\

        B_0L_V^{0.26}(1+B_1L_V)^{0.25},& \mbox{ $n_S<2.5$}
\end{array}
\right \}  \ {\rm kpc}
\label{shen2}
,\end{equation}
with $A_0=1.91$, $B_0=2.63$, $B_1=0.692$.
These parameters are valid assuming the concordance cosmological model 
($H_0=70$ km/s/Mpc, $\Omega _m=0.3$, $\Omega _\Lambda =0.7$).
For other cosmologies, we have to calibrate the relationship
between luminosities and radii with the corresponding luminosity
and angular distances for $z_{SDSS}\approx 0.1$. See Table \ref{Tab:shenpar}
for the values of $A_0$, $B_0$ and $B_1$ with other cosmologies. 
This relationship of Shen {\it et al.}\cite{She03} is fitted with galaxies of 
$-19>M_r>-24$ for early types and $-16>M_r>-24$ for late types, 
which is equivalent to
$2.5\times 10^9<L_V<2.5\times 10^{11}$ L$_{\odot,V}$ for early types
and $1.6\times 10^8<L_V<2.5\times 10^{11}$ L$_{\odot,V}$ for late types.
Our galaxies are within these limits.

In order to derive the luminosity, $L_{V,rest}$, for a given redshift ($z$) and the 
rest flux for the filter V, $F_{V,rest}$, 
we need the distance luminosity $d_L(z)$ from different 
cosmologies, such that (without considering neither evolution nor extinction)

\begin{equation}
d_L\equiv \sqrt{\frac{L_{V,rest}}{4\pi F_{V,rest}}}
\label{lrest}
.\end{equation}

And the average equivalent angular size evolution with $z$ should be
compared with the prediction of the same cosmology, which, in principle,
without taking into account the evolution, should be given by:

\begin{equation}
d_A(z)\equiv \frac{R_*}{\theta _{*,pred}(z)}
\label{da}
,\end{equation}
where $d_A(z)$ is the angular distance, and $R_*$ is the equivalent
physical radius of the galaxy associated with the equivalent
angular size $\theta _*$; that is, if the galaxy were  early-type
of V-rest luminosity $10^{10}$ L$_{\odot, V}$.

\subsection{Different cosmological scenarios}
\label{.cosmomodels}

\begin{enumerate}

\item Concordance model with Hubble constant $H_0=70$ km/s/Mpc, 
$\Omega _m=0.3$, $\Omega _\Lambda =0.7$:

\begin{equation}
d_A(z)=\frac{c}{H_0(1+z)}
\int _0^z\frac{dx}{\sqrt{\Omega _m(1+x)^3
+\Omega _\Lambda }}
,\label{concordance}\end{equation}
\begin{equation}
d_L(z)=(1+z)^2d_A(z)
\label{concordance2}
.\end{equation}

\item Einstein--de Sitter model [Eq. (\ref{concordance}) with $\Omega
_\Lambda =0$, $\Omega _m=1$]:

\begin{equation}
d_A(z)=\frac{2c}{H_0(1+z)}\left[1-\frac{1}{\sqrt{1+z}}\right]
,\end{equation}
\begin{equation}
d_L(z)=(1+z)^2d_A(z)
.\end{equation}

Although this is not the standard model nowadays, there are
some researchers who still consider it more appropriate than the concordance
model (e.g., \cite{Vau03,Bla06,And06}).
About the compatibility with type Ia supernovae data, see discussion in \S \ref{.hubble}.

Since the distances for objects at $z=0.1$ are 0.943 times the distances 
of the concordance model, the corrected relationship between
luminosity and radius is Eq. (\ref{shen2}) with $A_0=1.90$,
$B_0=2.61$, $B_1=0.699$.

\item Friedmann model of negative curvature with $\Omega =0.3$,  
$\Omega _\Lambda=0$, which implies a term of curvature $\Omega _K=0.7$
[\cite{Nab08}, Eq. (26)].

\begin{equation}
d_A(z)=\frac{c}{H_0(1+z)\sqrt{\Omega _K}}
\sinh \left(\sqrt{\Omega _K}\int _0^z\frac{dx}{\sqrt{\Omega _m(1+x)^3
+\Omega _K(1+x)^2 }}\right)
,\end{equation}
\begin{equation}
d_L(z)=(1+z)^2d_A(z)
.\end{equation}
I will use this model of an open universe to check that a model different 
to a flat universe does not change the results significantly.
Since the distances for objects at $z=0.1$ are 0.970 times the distances 
of the concordance model, the relationship between
luminosity and radius with corrected calibration  is
Eq. (\ref{shen2}) with $A_0=1.93$,
$B_0=2.59$, $B_1=0.736$.

\item Quasi-Steady State Cosmology (QSSC), 
$\Omega _m=1.27$, $\Omega _\Lambda =-0.09$, $\Omega _c=-0.18$ 
(C-field density) \cite{Ban99}

\begin{equation}
d_A(z)=\frac{c}{H_0(1+z)}
\int _0^z\frac{dx}{\sqrt{\Omega _c(1+x)^4+\Omega _m(1+x)^3
+\Omega _\Lambda }}
,\end{equation}
\begin{equation}
d_L(z)=(1+z)^2d_A(z)
.\end{equation}

This cosmology is not the standard model, but it can also fit
 many data on angular size tests\cite{Ban99}
or Hubble diagrams for SNe Ia\cite{Ban00,Nar02,Vis02}.
The expansion with an oscillatory term gives a dependence
of the luminosity and angular distance similar to the standard model,
adding the effect of matter creation (C-field) with 
slight changes depending on the parameters.
The parameters of this cosmology are not as well constrained as
those in the standard model. Here, I use the
best fit for a flat ($K=0$) cosmology given by Banerjee {\it et al.}\cite{Ban99}:
$\Omega _m=1.27$, $\Omega _\Lambda =-0.09$, $\Omega _c=-0.18$, which
corresponds to $\eta =0.887$ (amplitude of the oscillation relative to 1), 
$x_0=0.797$ (ratio between actual size of the Universe and the average
size in the present oscillation) and maximum allowed redshift of a galaxy
$z_{max}=6.05$ (note however that the maximum observed redshift has 
risen above 8 nowadays according to some authors, \cite{Tan09}). Other 
preferred sets of parameters give results that are close.
The  values most used are $K=0$, $\Omega _\Lambda =-0.36$, $\eta =0.811$,
$z_{max}=5$ \cite{Ban00,Nar02,Vis02,Nar07}, 
which imply $\Omega _m=1.63$, $\Omega _c=-0.27$,
but I avoid them because they do not allow  galaxies to be fitted with $z>5$.
Parameters with a curvature different from zero ($K\ne 0$) also give results
that are very close in the angular size test\cite{Ban99}.

Since the distances for objects at $z=0.1$ are 0.943 times the distances 
of the concordance model, the relationship between
luminosity and radius with corrected calibration is
Eq. (\ref{shen2}) with $A_0=1.94$,
$B_0=2.56$, $B_1=0.778$.

\item Static euclidean model with linear Hubble law for all redshifts:

\begin{equation}
d_A(z)=\frac{c}{H_0}z
\label{angdistst}
,\end{equation}
\begin{equation}
d_L(z)=\sqrt{1+z}d_A(z)
.\end{equation}

These simple relations 
indicate that  redshift is always proportional
to  angular distance, a Hubble law. We assume in this scenario that the
Universe is static; the factor $\sqrt{1+z}$ in the
luminosity distance stems from the loss of energy due to  redshift
without expansion. There is no time dilation, and precisely because
of that the factor is not $(1+z)$.
The caveat is to explain
the mechanism different from the expansion/Doppler effect, which gives rise
to the redshift. This cosmological model is not a solution which
has been explored theoretically/mathematically.
However, from a phenomenological point of view, we can 
consider this relationship between distance and redshift
as an ad hoc extrapolation from the observed dependence on the low redshift
Universe. Our goal here is to see how
well  it fits the data, and forget for the moment the theoretical
derivation of this law.

Since the luminosity distance for objects at $z=0.1$ is 0.966 times 
the luminosity distance of the concordance model 
and the angular distance is 1.115 times the angular distance of
the concordance model, we must adopt approximately Eq. (\ref{shen2}) 
with $A_0=2.23$, $B_0=2.98$, $B_1=0.741$.

\item Tired-light/simple static euclidean model :

\begin{equation}
d_A(z)=\frac{c}{H_0}{\rm ln} (1+z)
\label{angdisttl}
,\end{equation}
\begin{equation}
d_L(z)=\sqrt{1+z}d_A(z)
.\end{equation}

This is again a possible ad hoc phenomenological representation
which stems from considering that the photons
lose energy along their paths due to some interaction, and the
relative loss of energy is proportional to the length of that path
(e.g. \cite{LaV86}), i.e.
\begin{equation}
\frac{dE}{dr}=-\frac{H_0}{c}E
.\end{equation}
Of course, as in the previous case, this ansatz is very far
from being considered as the correct one by most cosmologists, 
but it is interesting to analyze its compatibility with the angular 
size test too.

For the calibration of Eq. (\ref{shen2}), I use the fact that
the luminosity distance for objects at $z=0.1$ is 0.921 times 
the luminosity distance of the concordance model, 
and the angular distance is 1.063 times the angular distance of
the concordance model, so $A_0=2.26$, $B_0=2.92$, $B_1=0.816$.

\item Tired-light/``Plasma redshift'' static euclidean model:

\begin{equation}
d_A(z)=\frac{c}{H_0}{\rm ln} (1+z)
,\end{equation}
\begin{equation}
d_L(z)=(1+z)^{3/2}d_A(z)
.\end{equation}

The plasma redshift application\cite{Bry04a}(\S 5.8) 
used $d_L(z)=(1+z)^{3/2}d_A(z)$ instead of 
$d_L(z)=(1+z)^{1/2}d_A(z)$ to take into
account an extra Compton scattering which is double that of the plasma redshift
absorption. For the calibration of Eq. (\ref{shen2}):
$A_0=2.00$, $B_0=2.78$, $B_1=0.674$.

\end{enumerate}

The volume element in a static uiverse is different from 
the volume element in the standard concordance model. 
Particularly for the standard concordance model the comoving volume element in a
solid angle $d\omega $ and redshift interval $dz$ is (for null curvature, which is
the case of the concordance model)
\begin{equation}
dV_{\rm concordance}=\left(\frac{c}{H_0}\right)^3
\frac{\left[\int _0^z\frac{dx}{\sqrt{\Omega _m(1+x)^3
+\Omega _\Lambda }}\right]^2}{\sqrt{\Omega _m(1+z)^3
+\Omega _\Lambda }}d\omega dz
\label{vol1}
\end{equation}
while for the two first static models it is
\begin{equation}
dV_{\rm static-lin.Hub.}=\left(\frac{c}{H_0}\right)^3z^2d\omega dz
,\end{equation}
\begin{equation}
dV_{\rm static-tir.l.}=\left(\frac{c}{H_0}\right)^3\frac{[{\rm ln}(1+z)]^2}
{(1+z)}d\omega dz
\label{vol2}
.\end{equation}
Hence, if we wanted to evaluate the evolution of some quantity
per unit comoving volume for the static universes, we must multiply
the result in the concordance model by the factor $\frac{dV_{\rm concordance}}
{dV_{\rm static}}$.

\subsection{Results}
\label{.results}

\begin{figure}
\vspace{1cm}
{\par\centering \resizebox*{8cm}{8cm}{\includegraphics{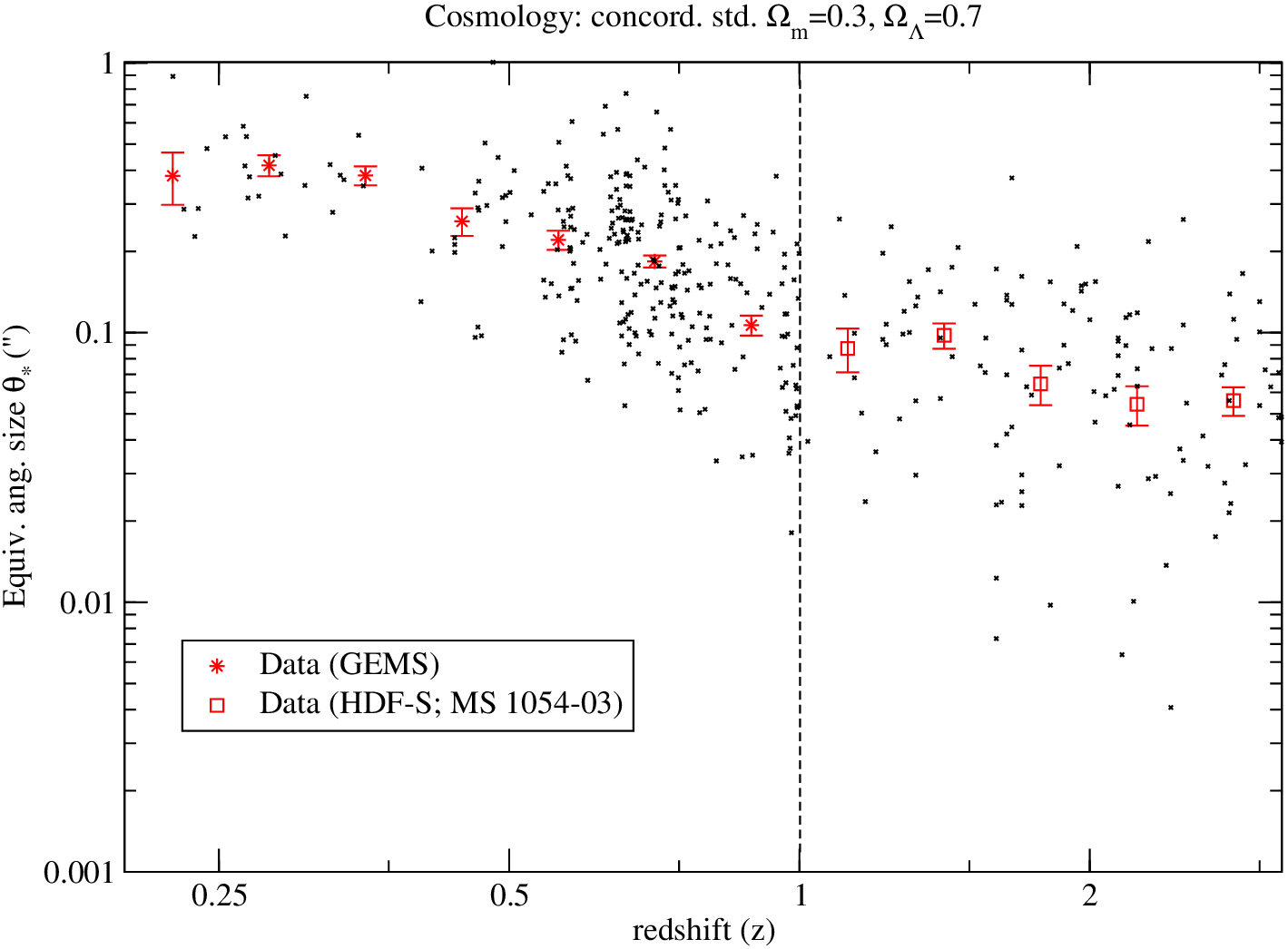}}}
\caption{Log--log plot of the equivalent angular size ($\theta _*$)
vs. redshift ($z$) in the concordance cosmology. Left: ($z<1$) 
GEMS data. Right: ($z>1$) HDF-S and MS 1054-03 data. 
The average of $\log _{10} \theta _*$ in bins of $\Delta \log _{10}(z)=0.10$
is represented with asterisks and squares with statistical error bars.}
\label{Fig:sizes0}
\end{figure}

\begin{figure*}[htb]
\vspace{1cm}
{\par\centering \resizebox*{4cm}{4cm}{\includegraphics{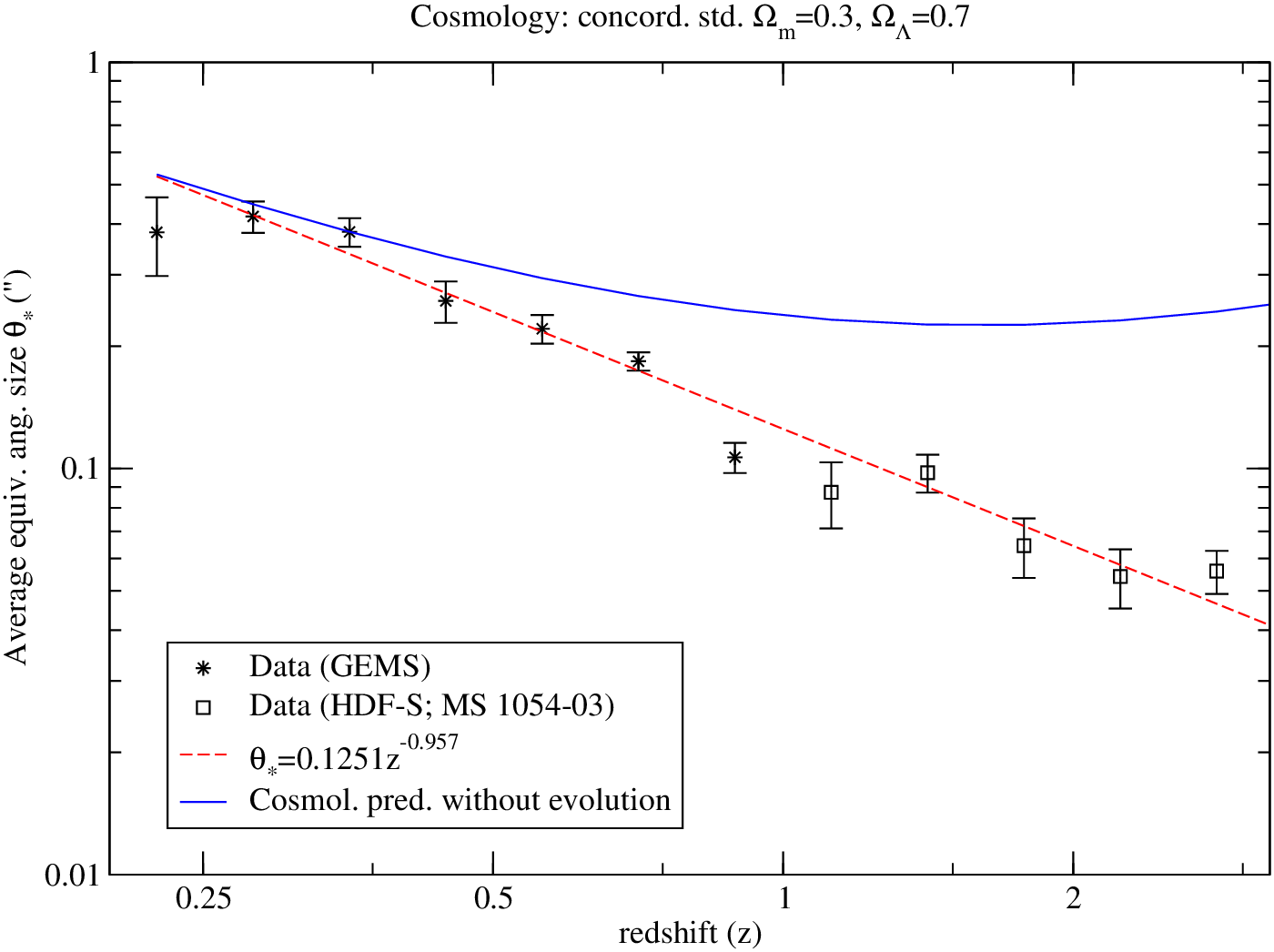}}
\hspace{1cm}\resizebox*{4cm}{4cm}{\includegraphics{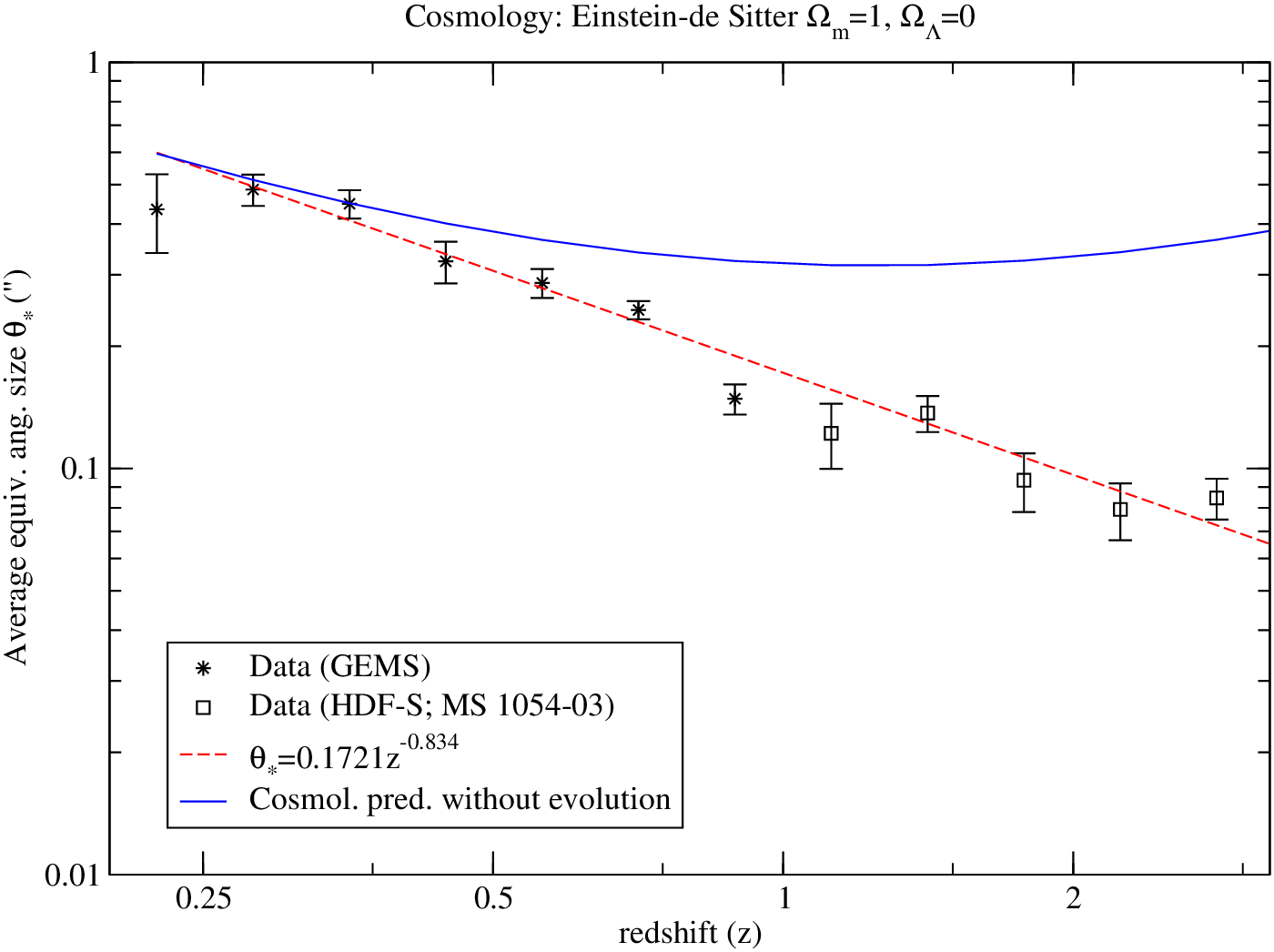}}\par}
\vspace{1cm}
{\par\centering \resizebox*{4cm}{4cm}{\includegraphics{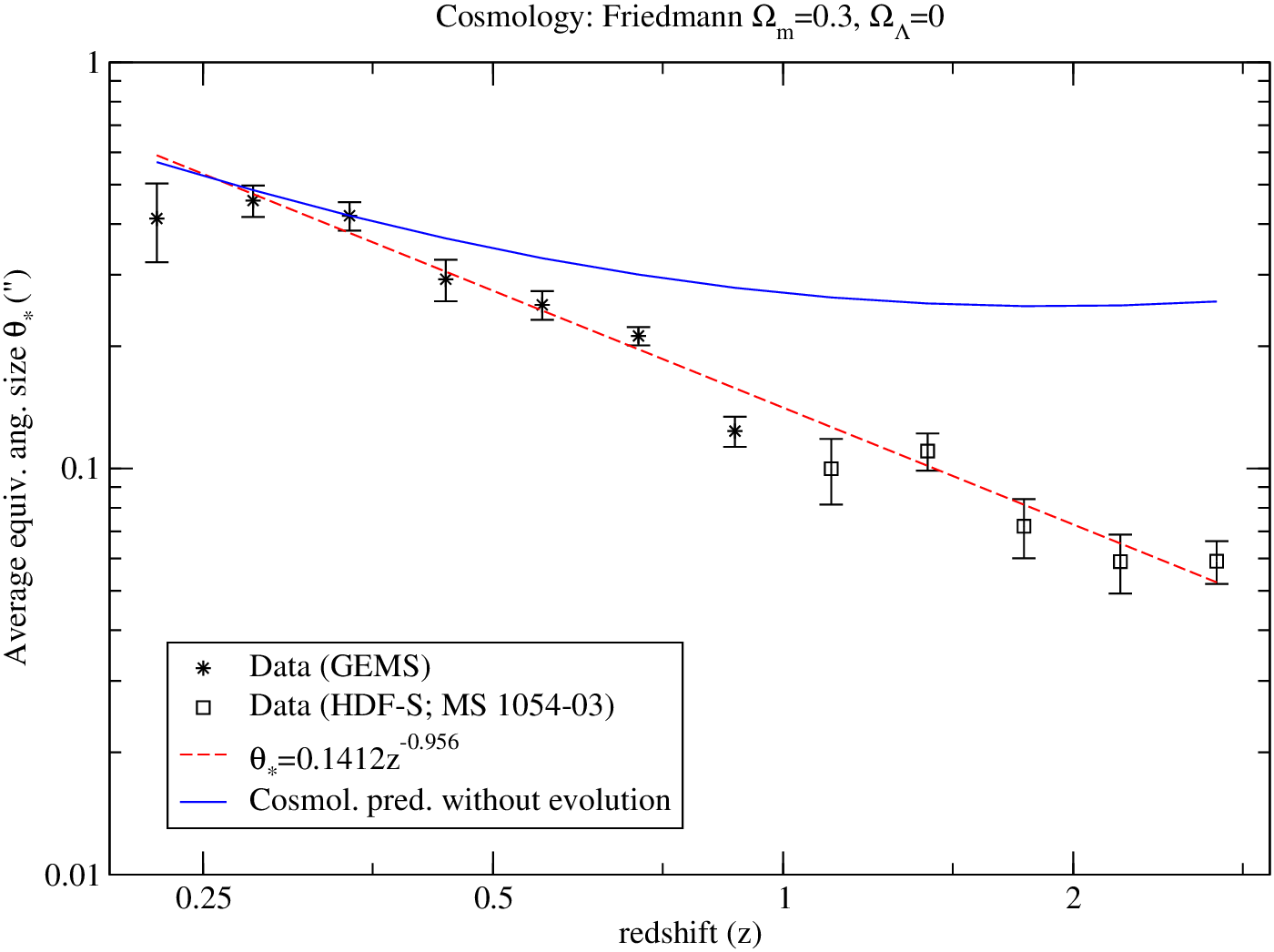}}
\hspace{1cm}\resizebox*{4cm}{4cm}{\includegraphics{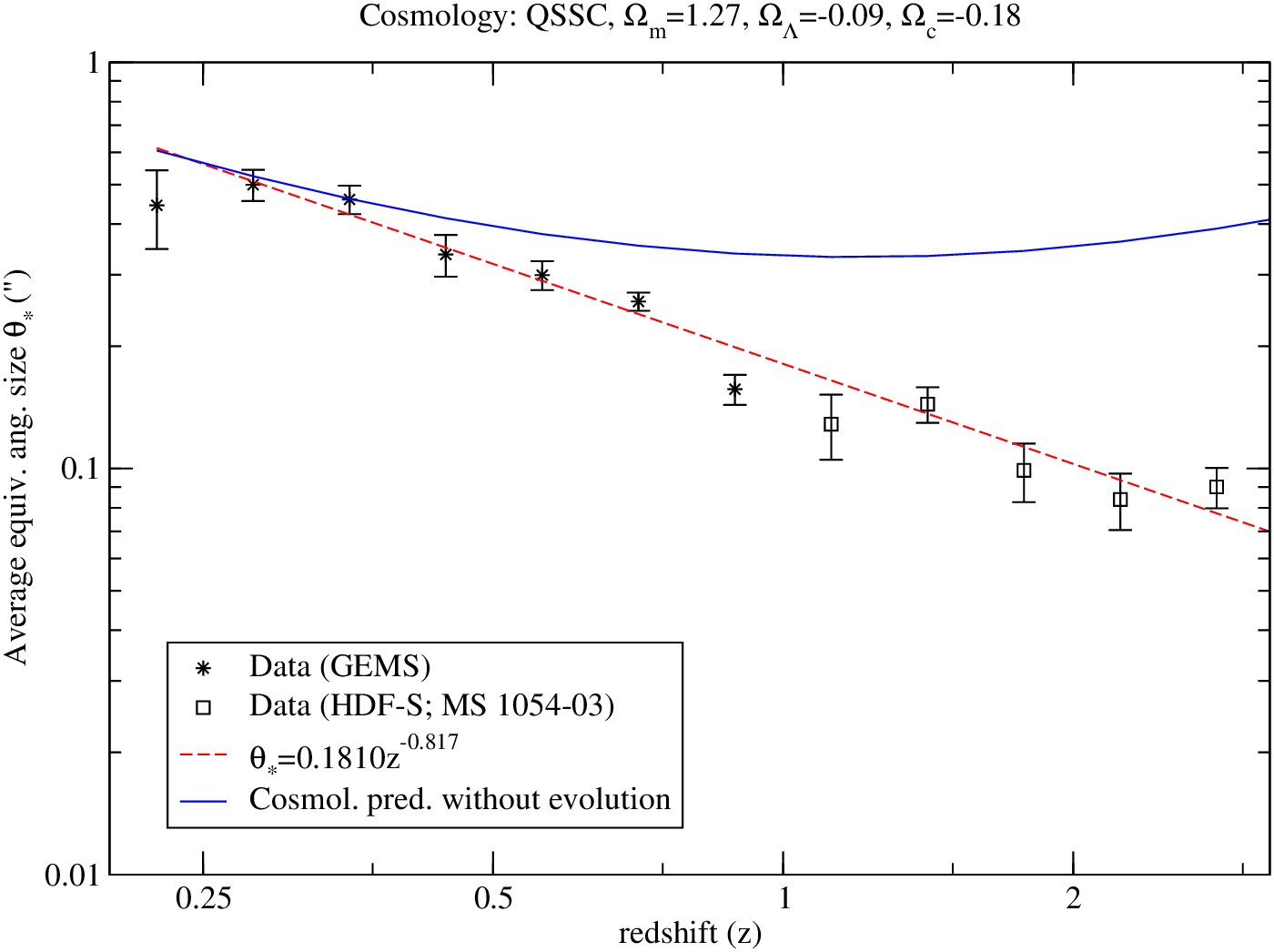}}\par}
\vspace{1cm}
{\par\centering \resizebox*{4cm}{4cm}{\includegraphics{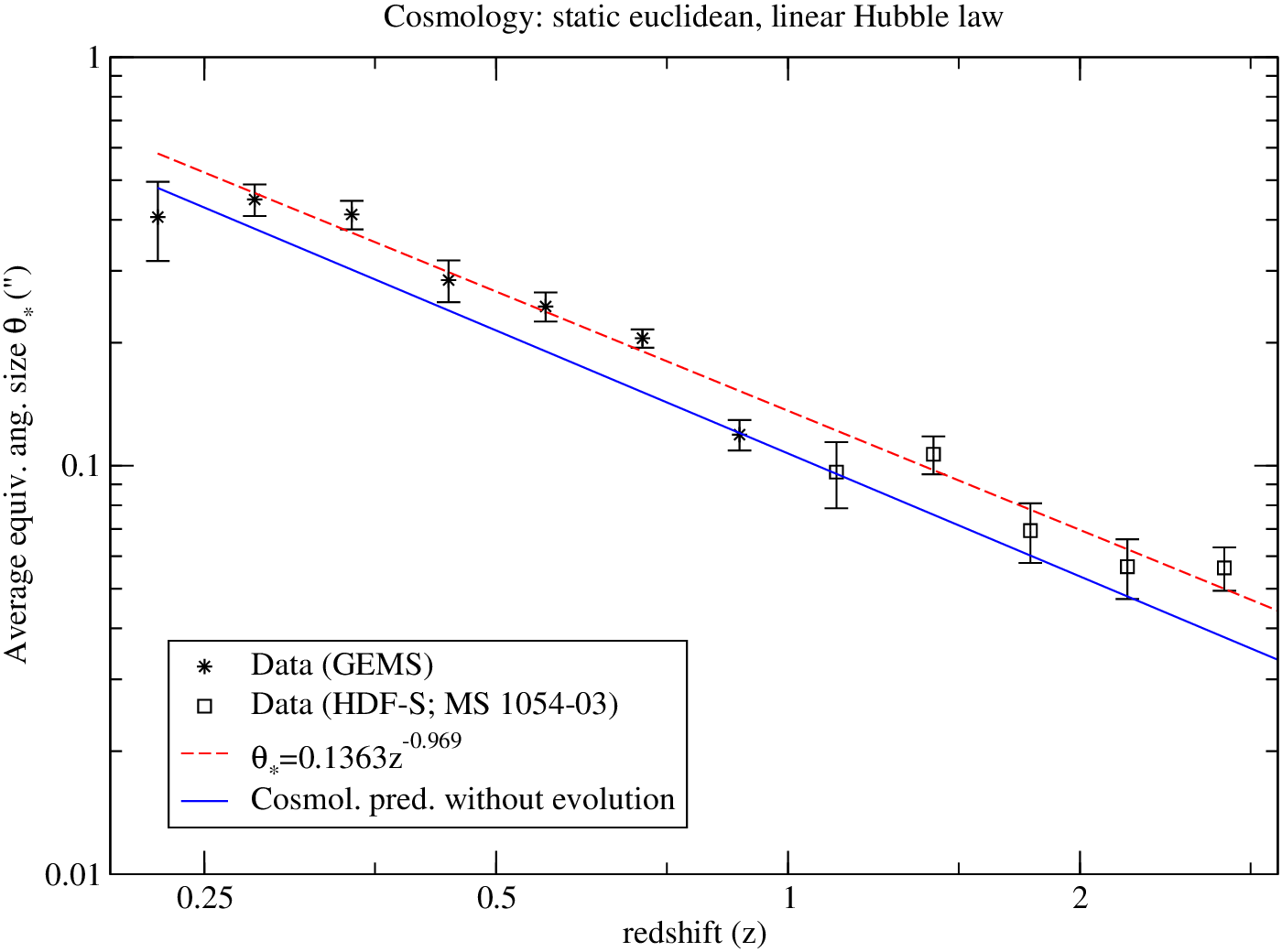}}
\hspace{.2cm}\resizebox*{4cm}{4cm}{\includegraphics{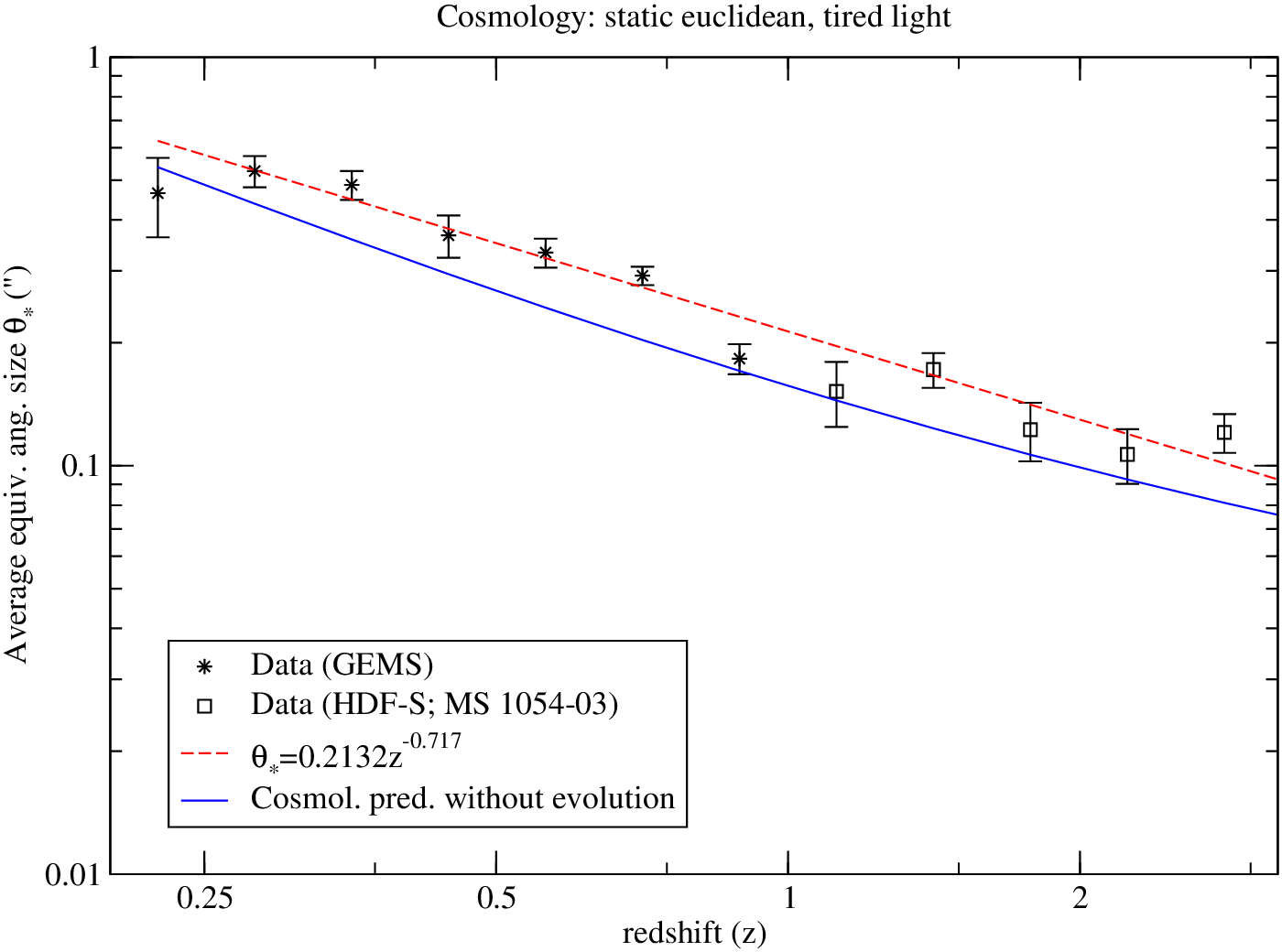}}
\hspace{.2cm}\resizebox*{4cm}{4cm}{\includegraphics{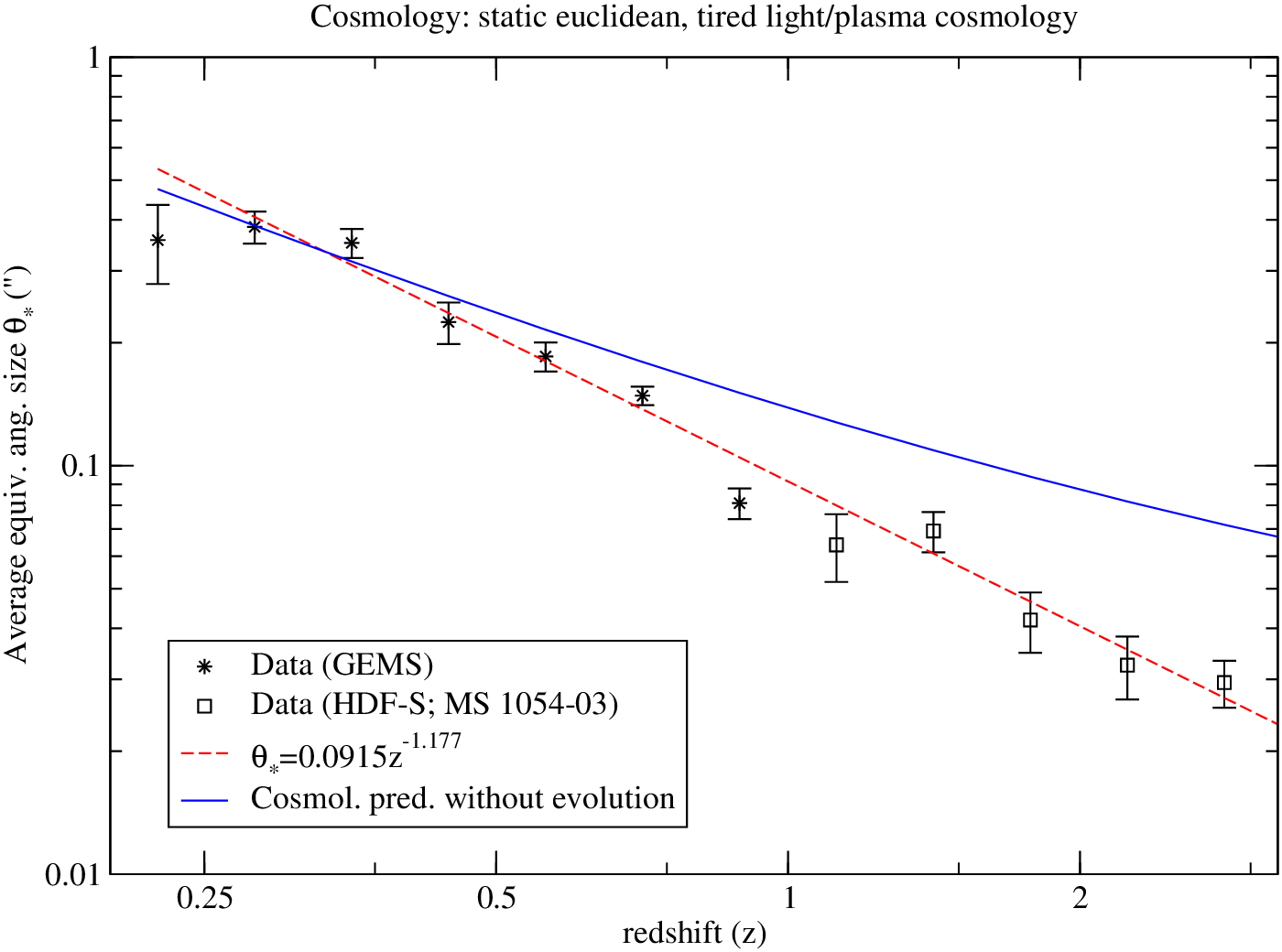}}\par}
\caption{}
Log--log plot of the average of $\log _{10} \theta _*$, where
$\theta _*$ is the equivalent angular size, vs. redshift ($z$). 
Bins of $\Delta \log _{10}(z)=0.10$. Error bars only represent
statistical errors; for the systematic errors, see text in \S 
\ref{.select}.
The seven plots are for the seven different cosmologies
described in \S \protect{\ref{.cosmomodels}}. Solid lines are
the model predictions (the normalization stems from the 
Shen {\it et al.}\cite{She03} calibration with
the corresponding parameters $A_0$, $B_0$ and $B_1$ in each 
cosmology). Dashed lines are the best weighted linear fits.
\label{Fig:angsize}
\end{figure*}

The results of the test are plotted in Fig. \ref{Fig:sizes0} for the
concordance model, with the equivalent angular size of each individual
galaxy, and Fig. \ref{Fig:angsize} for the different models, 
with the representation of the average value of 
$\log _{10}\theta _*$ in bins of $\Delta \log _{10}(z)=0.10$.
I do a weighted linear fit in the log--log plot. 
Since I am calculating an average of the logarithm for the angular
sizes, the possible error in individual galaxies should not significantly affect
the value of the average. 
The error bars in each bin of the plot are the statistical errors.
Note that there are values of $\theta _*$ lower than 0.03$''$, but they do indeed
correspond to measured values of $\theta \ge 0.03''$; 
$\theta _*$ is lower than $\theta $ at the highest redshifts because
the luminosity in V-filter of those galaxies is higher than $10^{10}$ L$_{\odot ,V}$.

As can be observed, the average
equivalent angular size gives a good fit to a $\theta _*(z)=Kz^{-\alpha }$ law,
with values of $K$, $\alpha $ given in Table \ref{Tab:fit}.
No fit is totally in agreement with the cosmological 
prediction without evolution and extinction.  
The static model with a linear Hubble
law is not very far from being compatible
with the data: I get $\theta _*(z)=0.136z^{-0.97}$,
near the expected dependence ($\theta _*(z)=0.109z^{-1}$)
although with a slightly larger size.

{\it {\Large \bf NOTE}: 
the normalization of the angular size in any model
prediction (solid line in Fig. \ref{Fig:angsize})
stems from the Shen {\it et al.}\cite{She03} calibration with
the corresponding parameters $A_0$, $B_0$ and $B_1$ in each 
cosmology. It does not stem from a fit.}

\begin{figure*}
\vspace{1cm}
{\par\centering \resizebox*{5cm}{5cm}{\includegraphics{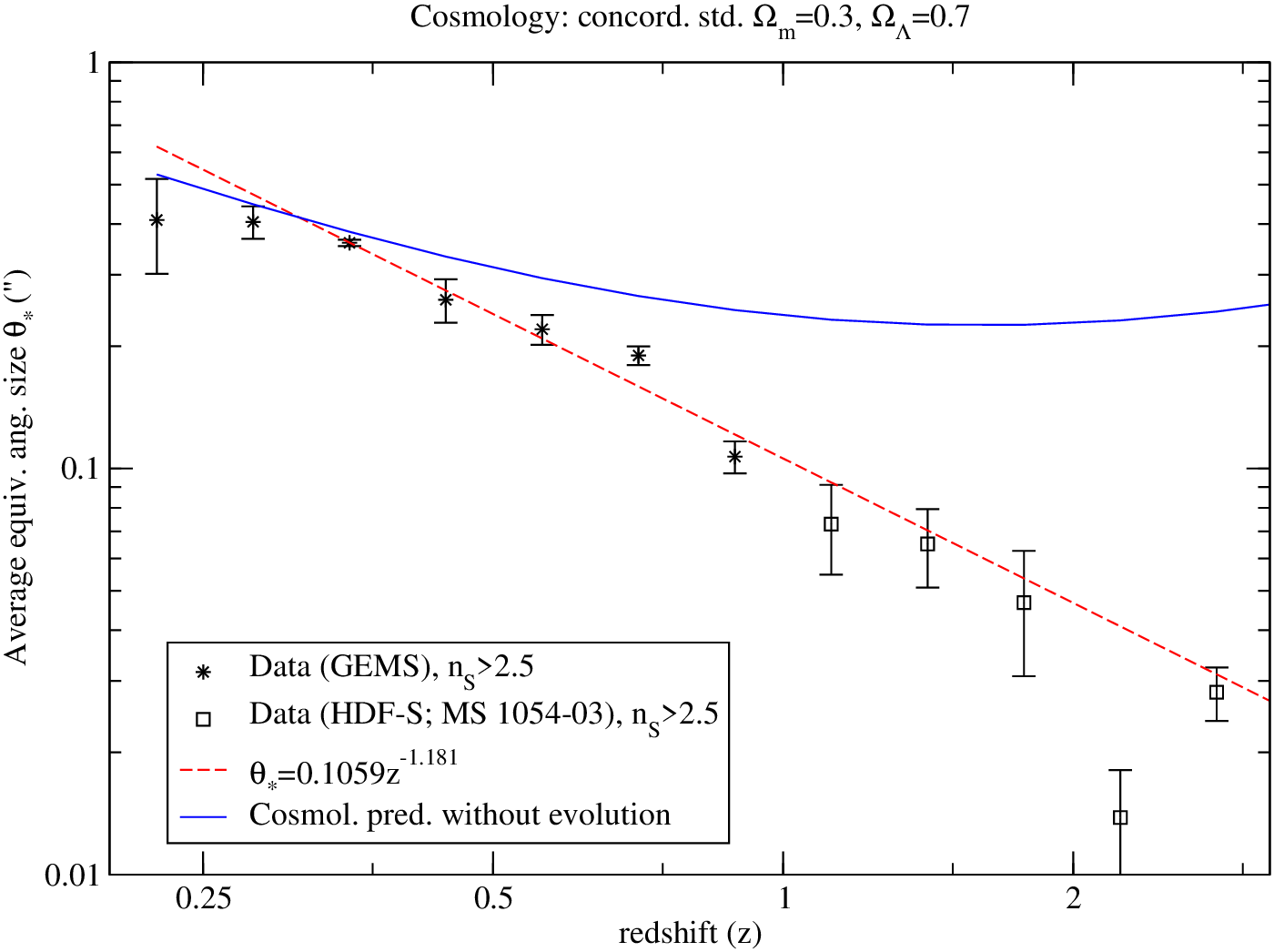}}
\hspace{1cm}\resizebox*{5cm}{5cm}{\includegraphics{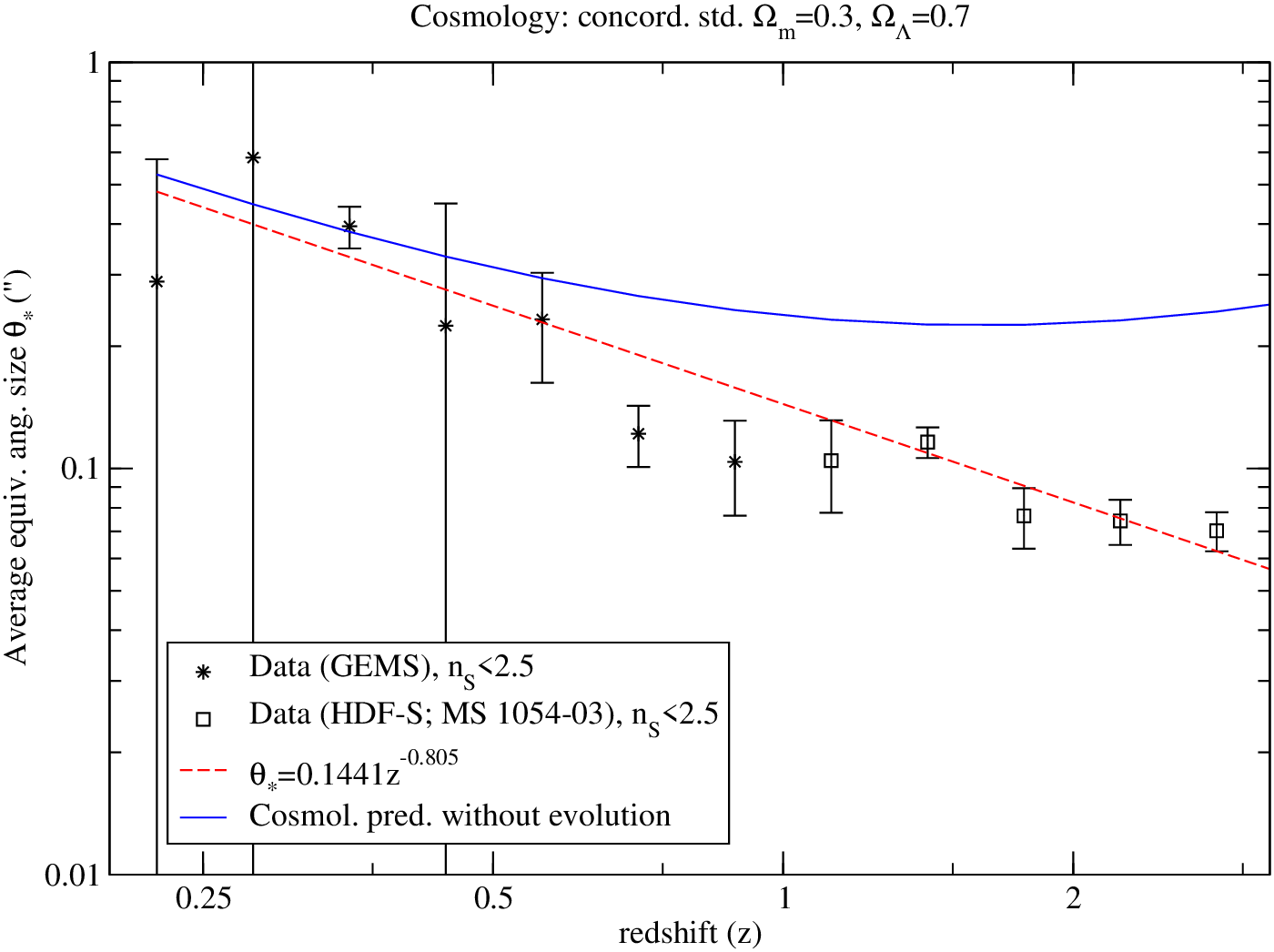}}\par}
\caption{Log--log plot of the average of $\log _{10} \theta _*$, where
$\theta _*$ is the equivalent angular size, vs. redshift ($z$). 
Bins of $\Delta \log _{10}(z)=0.10$.
The two plots are for the concordance cosmology separating 
elliptical galaxies ($n_S>2.5$) from disk galaxies 
($n_S\le 2.5$).}
\label{Fig:angsizetype}
\end{figure*}

If we separate elliptical ($n_S>2.5$) and disk galaxies 
($n_S<2.5$) for the concordance model,
the plots in Fig. \ref{Fig:angsizetype} are obtained, with
a higher slope for elliptical galaxies ($\alpha =1.18\pm 0.04$) than for
disk galaxies ($\alpha =0.80\pm 0.07$). This might be due to a different
evolution of disk and elliptical galaxies, but they
are likely to be due to different systematic errors for elliptical galaxies and
disk galaxies due to systematic errors in the value of
$n_S$. The galaxies with $2.0<n_S<2.5$
are suspected of being strongly contaminated by ellipticals since
the obtained $n_S$ tends to be lower than the real one 
(\cite{Tru06}, Fig. 1). If we take disk galaxies only
with $n_S<2.0$, then $\alpha =0.92\pm 0.06$, closer to unity. 
On the other hand, elliptical galaxies
with $\theta <0.125"$ are also strongly contaminated by
disk galaxies (\cite{Tru06}, Fig. 2) which are compacted by a wrong measure
of $n_S$ and consequently give a smaller radius than the real one.
If we take elliptical ($n_S>2.5$) galaxies only with $\theta >0.125"$:
$\alpha =1.03\pm 0.05$. Therefore, from the present
analysis and within the systematic errors, we cannot be
sure that the angular size test gives  different results
for elliptical and disk galaxies. However, when
all the elliptical and disk galaxies are put together, the excesses
and deficits more or less compensate; there are approximately 
as many elliptical galaxies misclassified as disk galaxies as
disk galaxies misclassified as elliptical galaxies.
See further discussion on the systematic errors in \S \ref{.select}. 

\begin{figure*}
\vspace{1cm}
{\par\centering \resizebox*{5cm}{5cm}{\includegraphics{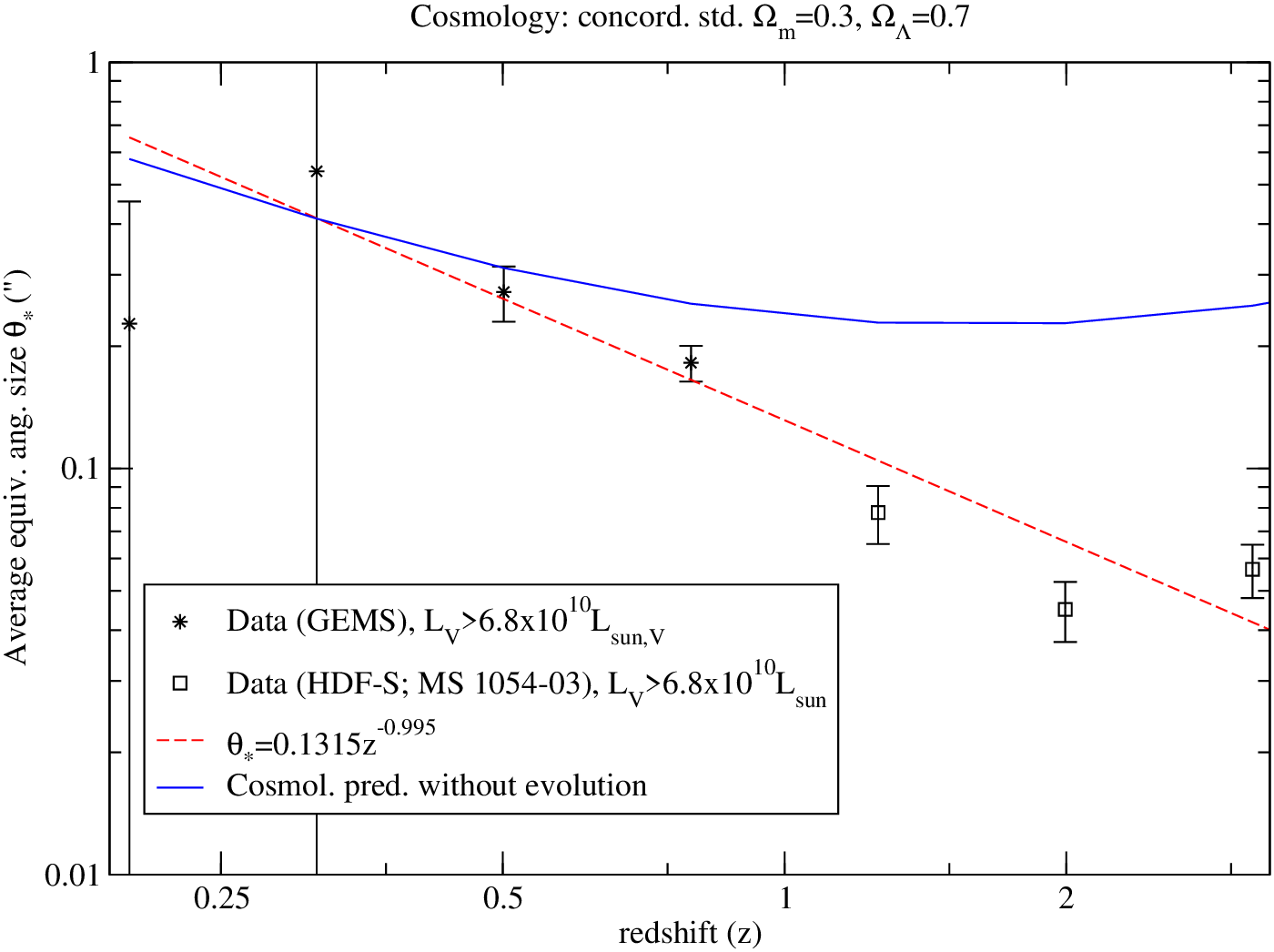}}
\hspace{1cm}\resizebox*{5cm}{5cm}{\includegraphics{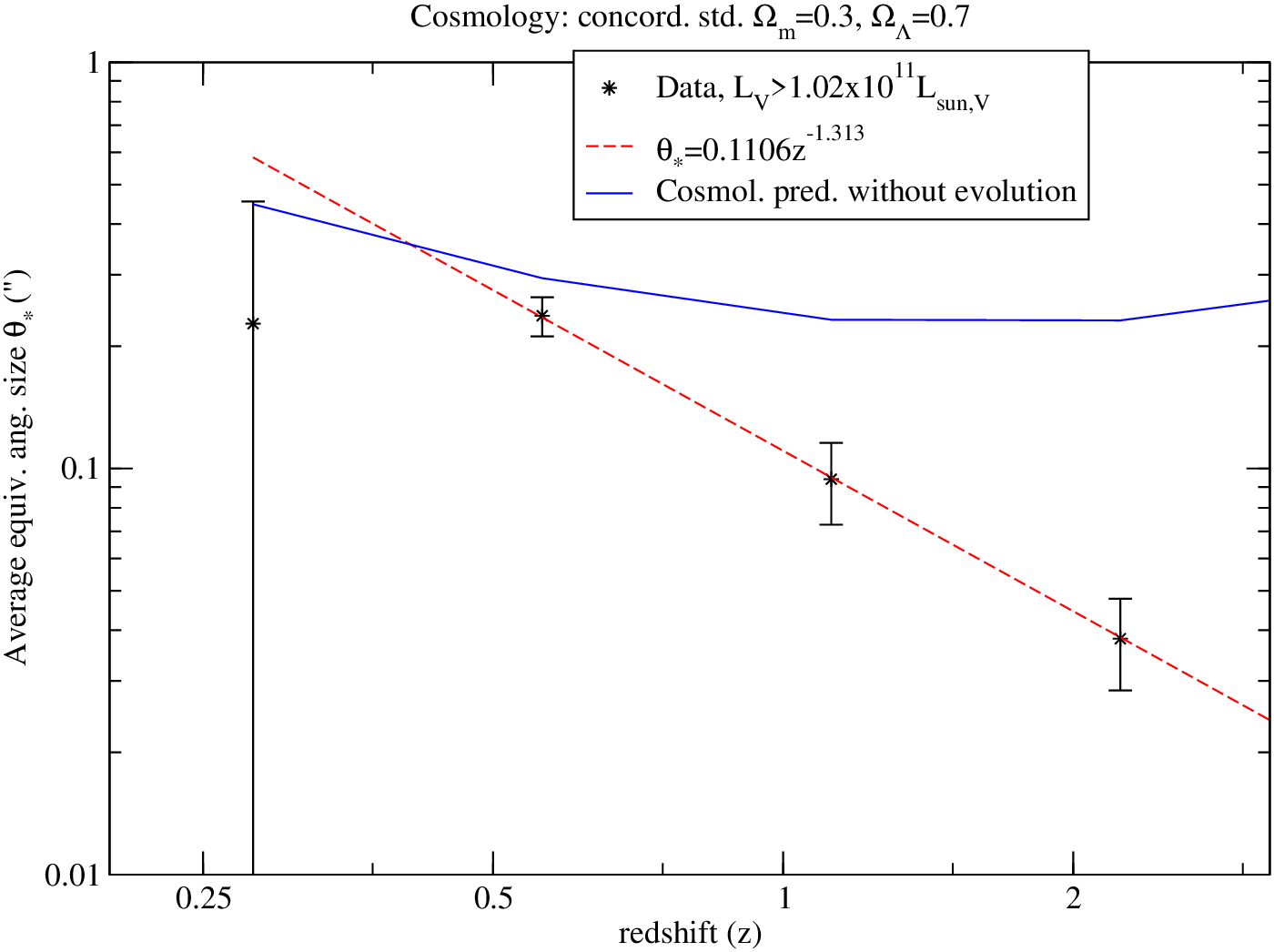}}\par}
\caption{Log--log plot of the average of $\log _{10} \theta _*$, where
$\theta _*$ is the equivalent angular size, vs. redshift ($z$). 
Bins of $\Delta \log _{10}(z)=0.20$ and $\Delta \log _{10}(z)=0.30$
respectively.
The two plots are for the concordance cosmology only for
galaxies with $L_V>6.8\times 10^{10}$ L$_{\odot ,V}$ (122 galaxies)
and $L_V>1.02\times 10^{11}$ L$_{\odot ,V}$ (48 galaxies)
respectively.}
\label{Fig:angsizelum}
\end{figure*}

In Fig. \ref{Fig:angsizelum}, I analyze the dependence of
$\alpha $ on the luminosity of the galaxies, and we see that
there is no significant dependence. $\alpha =1$ is
more or less compatible with all the subsamples of different
luminosity.

\begin{table}
\caption{}
Values of the parameters $A_0$, $B_0$ and $B_1$ in the
relationship between radius and luminosity of a galaxy, Eq.
(\ref{shen2}), calibrated
with different cosmological models.
\begin{center}
\begin{tabular}{cccc}
\label{Tab:shenpar}
Cosmology &  $A_0$ & $B_0$ & $B_1$ \\ \hline
Concord. $\Omega _m=0.3$, $\Omega _\Lambda =0.7$ & 1.91 & 2.63 & 0.692  \\
Einstein-de Sitter & 1.94 & 2.56 & 0.778  \\
Friedmann $\Omega _m=0.3$ & 1.93 & 2.59 & 0.736  \\
QSSC $\Omega _m=1.27$, $\Omega _\Lambda =-0.09$, $\Omega _c =0.18$ & 1.94 & 2.56& 0.778  \\ 
Static linear Hubble law & 2.23 & 2.98 & 0.741  \\
Static, tired light/simple & 2.26 & 2.92 & 0.816  \\ 
Static, tired light/plasma & 2.00 & 2.78 & 0.674  \\ \hline
St. lin. Hub. law, ext. $a_V=1.6\times 10^{-4}$ Mpc$^{-1}$ & 2.21 & 3.03 & 0.696  \\
St. tired light, ext. $a_V=3.4\times 10^{-4}$ Mpc$^{-1}$ & 2.22 & 3.01 & 0.719  \\
\end{tabular}
\end{center}
\end{table}

\begin{table*}
\caption{}
Weighted fit of the average
equivalent angular size for the data in Fig. \ref{Fig:angsize} 
to a law $\theta _*(z)=Kz^{-\alpha }$. Error bars give only statistical
errors and do not include systematic errors as explained in \S 
\protect{\ref{.select}}.
\begin{center}
\begin{tabular}{ccc}
\label{Tab:fit}
Cosmology &  K & $\alpha $ \\ \hline
Concord. $\Omega _m=0.3$, $\Omega _\Lambda =0.7$ & $0.1251\pm 0.0041$ &
$0.957\pm 0.045$ \\
Einstein-de Sitter & $0.1721\pm 0.0054$  & $0.834\pm 0.044$ \\
Friedmann $\Omega _m=0.3$ & $0.1412\pm 0.0046$ & $0.956\pm 0.045$ \\
QSSC $\Omega _m=1.27$, $\Omega _\Lambda =-0.09$, $\Omega _c =0.18$ & $0.1810\pm 0.0057$ & $0.817\pm 0.044$ \\ 
Static linear Hubble law & $0.1363\pm 0.0044$  & $0.969\pm 0.045$ \\
Static, tired light/simple & $0.2132 \pm 0.0066$ & $0.717\pm 0.043$ \\
Static, tired light/plasma & $0.0915 \pm 0.0031$ & $1.177\pm 0.047$ \\ \hline 
Concord., early type ($n_S>2.5$) & $0.1059\pm 0.0044$ & $1.181\pm 0.043$ \\
Concord., late type ($n_S\le 2.5$) & $0.1441\pm 0.0070$ & $0.805\pm 0.067$ \\
Concord., early type ($n_S>2.5$), $\theta >0.125"$ & $0.1226\pm 0.0060$ & $1.031\pm 0.051$ \\
Concord., late type ($n_S\le 2.0$) & $0.1600\pm 0.0073$ & $0.915\pm 0.057$ \\
Concord., $L_V>6.8\times 10^{10}$ L$_{\odot ,V}$ & $0.1315\pm 0.0083$ & $0.995\pm 0.098$ \\ 
Concord., $L_V>1.02\times 10^{11}$ L$_{\odot ,V}$ & $0.1106\pm 0.0117$ & $1.313\pm 0.186$ \\ \hline
St. lin. Hub. law, ext. $a_V=1.6\times 10^{-4}$ Mpc$^{-1}$ & $0.1055\pm 0.0035$ & $1.113\pm 0.046$ \\
St. tired light, ext. $a_V=3.4\times 10^{-4}$ Mpc$^{-1}$ &  $0.1565\pm 0.0050$ &$0.801\pm 0.044$ \\
\end{tabular}
\end{center}
\end{table*}

\subsection{Selection effects, and errors in the angular size
measurement and luminosity-radius relationship}
\label{.select}

As said previously, the definition of $\theta _*$ avoids 
selection effects due to Malmquist bias,
since $\theta _*$ for a given redshift should be nearly 
independent on how luminous the galaxy is, at least on average.

Nonetheless, although $R_*=\theta _*d_A$ is 
independent of the luminosity and type of the galaxy
at $z=0$, its possible evolution might depend on both parameters and this 
would give different average $\theta _*$ when the 
selection of galaxies is different. 
In any case, whatever  the sample of galaxies is, if we
are going to interpret the factor 
between the cosmological prediction and the observed average $\theta _*$
as a product of size evolution, this factor would
reflect the average evolution of $R_*$ in that sample. 
In our case, we are observing the factors for the average population
with $3.4\times 10^{10}<L_{V,rest}(L_\odot )<2.5\times 10^{11}$. 
For the concordance model the value of $R_*$ at $z=3.2$ is on
average $\approx 6$ lower than $R_*$ at $z=0$. 
At low redshift we know that the 
relationship of Eq. (\ref{shen2}) is correct (hence, the average size $R_*$ 
should not depend on the luminosity of these galaxies). Therefore,
the ratio of sizes between high/low redshift objects
will depend only on the variation of Eq. (\ref{shen2}) in
galaxies selected at high redshift.
We can say that the average size of the selected galaxies at 
$z=3.2$ is $\approx 6$ times lower than the galaxies at
low redshift with the equivalent luminosity and S\'ersic profile.
In other words, if there is a (very strong) evolution in
the size of the galaxies for a fixed luminosity (or a variation in
luminosity for a fixed radius), 
the factor by which the galaxies are smaller
will depend on which galaxies are selected, but in any case this
factor will represent an average galaxy shrinking factor.
An average factor of 6 in size reduction will mean that there 
are galaxies with size reduction by a factor larger than 6, and other
galaxies with size reduction by a factor smaller than 6.

Trujillo {\it et al.}\cite{Tru06}(\S 4.3) discussed the robustness of the
``average'' angular size measurement, whether it is affected
by some other biases or selection effects. Their tests showed that the
results, as presented here, are more or less robust and 
the difference in the results if we apply some minor corrections due to
uncertainties or incompleteness in the radius distribution
of the sources are $\Delta (\langle\log _{10}R\rangle)\sim 0.1$,
thus $\Delta \theta _*/\theta _*\sim 20$\% .

There is a minimum angular size that Hubble telescope or FIRES can observe 
below which the uncertainties of $\theta $ and $n_S$
are very high, with important systematic deviations.  
Normally, the angular sizes are overestimated for these
low angular sizes, so the observed average $\theta _*$ 
increases with $z$ as more and more of the 
smaller angular size galaxies are included in the average.
Moreover, the very compact galaxies are not included in the
sample because they might be classified as stars. 
This affects mainly the $z>1$ data, where the limit for 
low systematic errors is around $\theta =0.125"$ \cite{Tru06}. 
Suppression of these galaxies
would change the average of $\theta _*$ at $z>1$ by 10--30\%.
Systematic errors for $\theta <0.125''$ are up to 50\% or lower (\cite{Tru06}, 
Fig. 2), which produces a systematic error
on the average $\theta _*$ lower than $\approx 15$\%.

Another effect to investigate is the error in the relationship of Eq.
(\ref{shen2}). This represents the average radius for a given luminosity
and there is some dispersion of values with respect to it (given
by Shen {\it et al.}\cite{She03}). I am not interested here in these statistical errors
of $R$ because this would just affect  a dispersion of
values of $\theta _*$ without changing its average value. Our concern
is about the systematic errors in Eq. (\ref{shen2}). I will
analyze five sources of systematic errors:

\begin{enumerate}

\item The color correction of $\langle (V-r'_{SDSS,AB})\rangle$=0.33 
for early-type galaxies and $=0.30$ for late-type galaxies \cite{Fuk95} 
might change if there is a preferential galactic
type within the group of early- or late-type galaxies is given for some redshift. 
Indeed, Fukugita {\it et al.}\cite{Fuk95} give a value of 
$\langle (V-r'_{SDSS,AB})\rangle$ of
0.36 for E type, 0.31 for S0, 0.32 for S$_{ab}$, 0.29 for S$_{bc}$,
0.28 for S$_{cd}$. That is, there are variations up to $\approx 0.03$
with respect to the average in early types, and 0.02 for late types. 
In the worst  cases, these systematic errors in colors would produce
a systematic error of $\approx 2$\% in $\theta _*$, which is negligible.

\item A misclassification of an elliptical galaxy as disk galaxy or vice versa
produces an error up to a factor 2 in the value of $R$ derived from
Eq. (\ref{shen2}), which leads to an error of factor $<2$ in $\theta _*$.
The effect of the uncertainty in $n_S$ for small angular sizes 
has already been checked by Trujillo {\it et al.}\cite{Tru06}(\S 4.3) and the error of the
average of $\theta _*$ is lower than $\sim 20$\% for the sample with 
all of the galaxies.
However, when I divide the galaxies into elliptical and disk types
(see \ref{Fig:angsizetype}) the effect might be larger, as 
was noted previously concerning Fig. \ref{Fig:angsizetype}.

\item Systematic errors in the luminosity at V-rest due to 
errors in the photometric redshift amount less than 4\% \cite{Rud06}, 
which implies errors in $\theta _*$ less 
than 2.5\%. However, the systematic error of the luminosity 
measurement in faint elliptical galaxies amounts around 15\% \cite{Tru06}, 
which translates into 6\% of error in $\theta _*$.

\item The calibration of eq. (\ref{shen2}) (calibration of
the parameters $A_0$, $B_0$ and $B_1$) is done 
for cosmologies different to concordance model 
assuming that the SDSS galaxies have redshift $z_{SDSS}=0.1$. There is indeed
a dispersion of redshift values  in SDSS galaxies around this average.
If we set $z_{SDSS}=0.15$ we would obtain, for instance for the
static linear Hubble law ansatz, $A_0=2.32$, $B_0=3.15$, $B_1=0.721$;
the variations in the average measured $\theta _*$ are negligible
($<<1$\%), but the predictions of the cosmological model for $\theta _*$
(proportional to $A_0$) increase by $\approx 6$\%. That is, the ratio
between measured and predicted $\theta _*$ decreases by 6\%.
An error of 0.05 in the average redshift is among the worst cases,
so we can say that the systematic error in the calibration should
be less than $\approx 5$\%.

\item Another systematic error comes from the $V_{max}$
correction of the selection effects in the SDSS data made by
Shen {\it et al.}\cite{She03}(\S 3.1). Shen {\it et al.} \cite{She03} 
make a correction of selection
effects by assigning a weight to each galaxy inversely proportional
to the maximum comoving volume within which galaxies identical to one under
consideration can be observed\cite{Qin97}. 
This correction mainly affects 
faint elliptical galaxies, as can be 
observed in fig. 2 of Shen {\it et al.}\cite{She03} and amounts 
to $\frac{\Delta R}{R}<\sim 0.7\times exp(-0.7L_V)$ 
in the average radius of the galaxies
for elliptical galaxies of luminosity $L_V$ (in units of 
$10^{10}$ L$_{\odot, V}$).

The weights depend on the cosmological
model, since the comoving volumes at a given redshift change in the
different cosmological models. Moreover, there is an intrinsic 
systematic error in using the standard model:
Shen {\it et al.}\cite{She03} used the integration of physical volumes instead
of comoving volumes [see Eq. (8) of Shen{\it et al.}\cite{She03}], which
leads to systematic errors of $V_{max}^*$ up to $\sim 50$\%
[there is a factor $(1+z)^3$ between comoving and physical volumes]. 
In order to calculate the effect of a change of cosmology, apart
from the recalibration of $A_0$, $B_0$ and $B_1$, I would need to have
all their SDSS data at hand and repeat the full analysis of their
fit with the new conditional maximum volume $V_{max}^*$ values for
each cosmology [see Eq. (9) of Shen {\it et al.}\cite{She03}] using the correct
comoving volume $V_{max}$ instead of the integration of physical
volumes, which is not possible for us. Assuming a
total systematic error in the volume of $\Delta V_{max}^*\sim 0.5V_{max}^*$,

\begin{equation}
\Delta (\langle\log _{10}R\rangle)<\sim 0.15\times exp(-0.7 \langle L_V\rangle)
.\end{equation}
Since $L_V>3.4$ in our selected sample, 
this may justify systematic errors in $R(L)$ up to 3\% .

\end{enumerate}

Summing up, apart from the statistical errors plotted in
Fig. \ref{Fig:angsize}, there are systematic errors in the average 
$\theta _*$ that may amount up to 30--40\% for the general
case with all the galaxies

\subsection{Relationship with the surface brightness test}

The average surface brightness of a galaxy with total flux
$F_{\lambda ,rest}$ and half-light circularized angular size $\theta $
is:

\begin{equation}
SB=\frac{F_{\lambda ,rest}/2}{\pi \theta ^2}
.\end{equation}
Using eqs. (\ref{lrest}), (\ref{da}); and $d_L=(1+z)^{i/2}d_A$,
with $i=1$ if static, and $i=4$ if expanding, it is

\begin{equation}
SB=\frac{L_{\lambda ,rest}}
{8\pi ^2R^2(1+z)^i}
\label{sb}
.\end{equation}

The intrinsic surface brightness [without the $(1+z)^i$ dimming factor] 
is
 
\begin{equation}
SB_0=\frac{L_{\lambda ,rest}}{8\pi ^2R^2}
\label{sb0}
.\end{equation}
Given the relationship between radius and luminosities of eq. 
(\ref{shen2}), we find that the average intrinsic surface
brightness in V-rest (assuming no size evolution) should follow:

\begin{equation}
SB_{0,V-rest}(L_V,n_S)=\left \{ \begin{array}{ll}
        \frac{L_V^{-0.30}}{8\pi ^2A_0^2},& \mbox{ $n_S>2.5$} \\

        \frac{L_V^{0.48}}{8\pi ^2B_0^2(1+B_1L_V)^{0.50}},& \mbox{ $n_S<2.5$}
\end{array}
\right \}  
.\end{equation}
The surface brightness for a given luminosity should
be independent of the redshift if there is no evolution in size.
I can also define an equivalent surface brightness to avoid the
dependence on the luminosity:

\begin{equation}
SB_*\equiv \frac{F_{V,rest}}{2\pi \theta ^2}
\frac{SB_0(10^{10}\ {\rm L_{\odot ,V}},>2.5)}
{SB_0(L_{V,rest},n_S)}=\frac{10^{10}\ {\rm L_{\odot ,V}}}
{8\pi ^2R_*^2(1+z)^i}
.\end{equation}

This is indeed the surface brightness test, or Tolman test\cite{Hub35}.
In Fig. \ref{Fig:angsize} for the static models, 
it is observed that the data of $\theta _*(z)$ without size
evolution are fitted more or less (within
the statistical+systematic error). Thus, 
$R_*$ is nearly constant in a static model for all redshifts and we obtain
$\langle SB_*\rangle \propto (1+z)^{-1}$, while for the expanding
model we would need a strong evolution in $R_*(z)$ and
the average surface brightness would be 
$\langle SB_*\rangle \propto R(z)^{-2}(1+z)^{-4}$. In order to give
the same $\langle SB_*\rangle (z)$ as in the static Universe,
assuming that $d_L(z)$ is approximately the same (which is
nearly true for the concordance and the linear Hubble law luminosity
distances; see \S \ref{.hubble}), $R(z)\approx R(z=0)(1+z)^{-3/2}$.

Lubin \& Sandage\cite{Lub01} obtained  
$\langle SB\rangle \propto (1+z)^i$ with $i=1.6-3.2$ for a
sample with $z<0.9$ depending on the filter 
(R or I), and the initial hypothesis (expanding or static). 
Lerner\cite{Ler06} obtained $i=1.03\pm 0.15$ for $z<\sim 5$, 
with data in wavelengths
from ultraviolet to visible from Hubble Space Telescope
(see also Lerner\cite{Ler09}),
while Nabokov \& Baryshev\cite{Nab08} obtained $i=3-4$ with the
same type of data but without including K-corrections.
Andrews\cite{And06} obtained $i=0.99\pm 0.38$, $i=1.15\pm 0.34$, 
with two different samples of cD galaxies.
Lubin \& Sandage argue that their data are compatible with 
an expanding Universe if we take evolution into account.
Lerner criticizes Lubin \& Sandage
for not using the same range of wavelengths at rest for the
high and low redshift galaxies but instead using K-corrections with
many free parameters, which is less direct and 
more susceptible to errors. Moreover, Lerner argue against the
evolution that the intrinsic ultraviolet surface brightness
of high redshift galaxies would be extremely large, with impossible
values (see \S \ref{.uv}).

\section{Evolution of galaxies in expansion models}
\label{.evol}

From the plot in Fig. \ref{Fig:angsize}(concord. model), we see that, in order
to make the concordance model compatible  with the data on angular size,
we must assume an evolution such that the galaxies at high redshift are
much smaller than at low redshift ($z<\sim 0.2$). For instance, at redshift
$z=2.5$, the galaxies should be on average $\approx 4.6$ times smaller.
Trujillo {\it et al.}\cite{Tru06} obtained a lower average factor ($\sim 3$) 
because they used only galaxies with the constraint 
$\theta >0.125''$; however in the error bars of Trujillo et al.
result,  this bias effect of avoiding $\theta \le 
0.125''$ galaxies is included, and within this error bar 
it is compatible with our result.  
The most massive galaxies are thought to be contracted by a factor
of $\approx 4$ up to $z=1.5$ \cite{Tru07} and
of $5.5$ up to $z=2.3$\cite{van08}, but Mancini {\it et al.}\cite{Man09} think that
the extra compactness of these galaxies can be understood in terms of fluctuations 
due to noise preventing the recovery of the extended low surface brightness halos 
in the light profile, so those factors might be not real. 
For galaxies with redshift $z=3.2$ in our Fig. 
\ref{Fig:angsize}(concord. model), 
the average ratio of sizes would increase to a factor 6.1
(comparing the linear fit with the theoretical prediction).
Separating by types, the ratio would be 4.5--6 for disk galaxies
and 6--10 for elliptical galaxies, taking into account the systematic
uncertainties commented in \S \ref{.results}.
Other authors\cite{Fer04,Bou04,Tru04,Wik08,Cim08,vand08,Nab08} 
 find similar results at high redshift.

If we have galaxies that are on average 6 times smaller than the 
equivalent galaxies (of same type and luminosity) at low redshift,
this means that the V-rest luminosity density (inversely proportional to the
cube of the size) of these galaxies is $\approx 200$ times higher 
than at low redshift. The surface brightness in V-rest is increased by a 
factor $\sim 40$. Is this situation possible?

\subsection{Effect of the expansion of the Universe}
\label{.expansion}

In the standard picture of the expanding models, the expansion does not
affect the galaxies because there is no local effect on particle dynamics 
from the global expansion of the universe: the tendency to separate 
is a kinematic initial condition, and once this is removed, all memory 
of the expansion is lost\cite{Pea08}.
Note, however, that there are other views on the topic of whether there is
expansion of galaxies due to the expansion of the Universe 
and  the nature of the expansion itself \cite{Fra07,Bar08}.
Dark energy or cosmological vacuum might have some influence
on the size of the galaxies \cite{Now01,Ser07}, but this effect would be small.
Lee\cite{Lee08} suggested instead that the 
size of the galaxies increases as the scale factor of the universe
assuming dark matter models based on a Bose--Einstein condensate or 
scalar field of ultra-light scalar particle, which is another
heterodox idea to explore. Nonetheless, apart from this
kind of proposals, within standard scenarios, the galaxies do not expand
with the Universe.

An indirect way in which the expansion has an effect is in the formation of galaxies.
In the theoretical $\Lambda $CDM hierarchical scenarios, galaxies formed 
at higher redshift should be denser\cite{Mo98} since, to decouple
from expansion, structures must have a given density ratio with the
surrounding density, which is larger at higher redshift. Some authors
(e.g., \cite{Tru06,Fer04}) have used this argument to explain the apparent
size evolution. 
However, the observed redshift in galaxies is not its formation redshift, 
so the application of this idea is not straightforward.
As a matter of fact, the stellar populations
of most local massive elliptical galaxies are very old\cite{Jim07}, 
and formed before the age corresponding to 
$z\sim 3$, so we cannot say that galaxies now are larger
because of this effect since most of them were formed $>10$ Gyr ago.
The difference of formation epoch of the galaxies 
observed now and at $z\sim 3$ is not large.
Moreover, the hierarchical scenario in which massive galaxies form first do
not represent the observed Universe appropriately, as I argue in \S \ref{.mergers}.

\begin{figure}
\vspace{1cm}
{\par\centering \resizebox*{8cm}{8cm}{\includegraphics{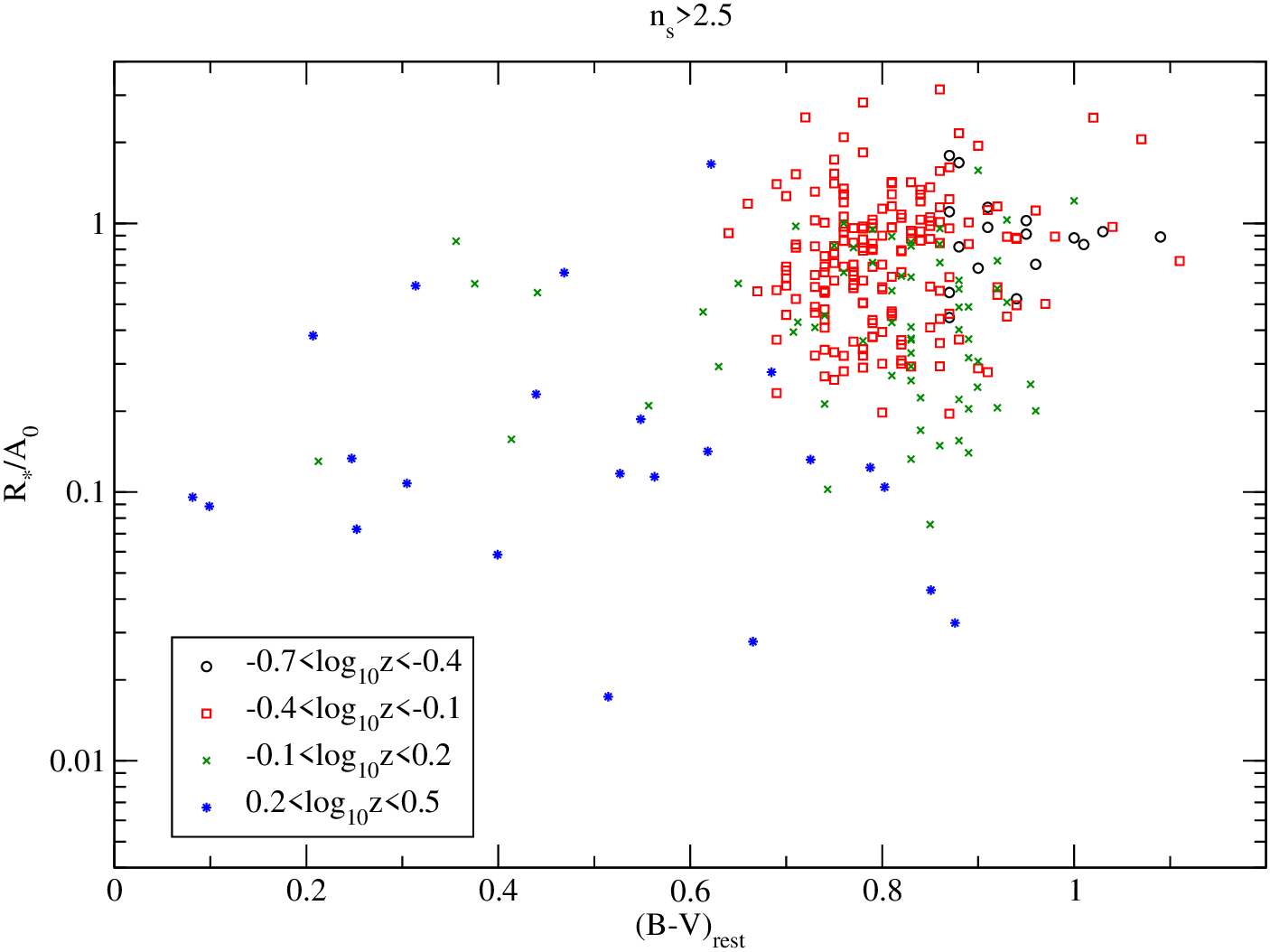}}}
\caption{}
Equivalent linear size (normalized with Shen {\it et al.}'s\cite{She03} calibration)
vs. $(B-V)$ color at rest for elliptical galaxies
in the concordance cosmology.
\label{Fig:bmv}
\end{figure}

Furthermore, the theoretical claim by Mo {\it et al.}\cite{Mo98} derived from the models 
that galaxies which formed earlier are denser is not in general observed.
If it were true, we should observe that at low redshift the youngest galaxies
(formed later) should be much larger for a given mass
than the oldest galaxies of age 12--13 Gyr. There
is already evidence that this is not the case: the densest galaxies are young
instead of old\cite{Tru09}. And we can check with our own sample within $0.2<z<3.2$ that
the color of elliptical galaxies is not correlated with size: Fig. \ref{Fig:bmv}.
Redder elliptical galaxies are older and, for a given redshift, indicate earlier
formation, which should be equivalent to smaller size, at least statistically.
This correlation is not observed at all: linear fits in the four redshift ranges
of Fig. \ref{Fig:bmv} all give slopes compatible with zero within $1\sigma$ except
for the range $-0.4<\log_{10}(z)<-0.1$, which gives
$\frac{d(log_{10}(R_*/A_0))}{d(B-V)_{\rm rest}}=+1.0\pm 0.5$,
a $2\sigma $ correlation but in a direction opposite to prediction that galaxies
are larger when redder=older (formed earlier).

Therefore, it is not a question of comparing 
the formation of galaxies at different redshifts but the evolution of galaxies 
already formed, either isolated or in interaction/merging with other systems.

\subsection{Younger population of high redshift galaxies}
\label{.brighter}

The main argument in favour of the evolution in size for a fixed
luminosity is that the younger a galaxy is the brighter it is, and
we expect to see younger galaxies at high redshift. Therefore,
galaxies with radius smaller than $R_*$ in the past will 
produce the same luminosity as galaxies with that radius at
present. How much brighter?

Using Vazdekis {\it et al.}'s\cite{Vaz96} synthesis model, 
with revised Kroupa IMF, we can derive the mass--luminosity ratio in a passively
evolving elliptical galaxy as a function of its intrinsic (B-V) color and its
luminosity. The metallicity degeneracy is
broken with an iterative method which uses the relationship between 
stellar mass and metallicity\cite{Lop09}. With this method,
the average mass--luminosity ratio of 
elliptical galaxies (not affected by extinction) at the last bin of
Fig. \ref{Fig:angsize} ($z=2.5-3.2$) is $\langle M_*/L_V\rangle =0.5$
[$N=6$, $\langle L_V\rangle =10^{11}$ L$_\odot $, $\langle (B-V)\rangle =0$]. Rudnick
{\it et al.}'s\cite{Rud06}(\S 4.4) analysis of the same galaxies
also obtain a quite similar mass--luminosity ratio.
For the elliptical galaxies with $\langle L_V\rangle >5\times 10^{10}$ L$_\odot $
of the two first bins ($z=0.20-0.32$): 
$\langle M_*/L_V\rangle =2.2$ [$N=7$, $\langle L_V\rangle =7\times 10^{10}$ 
L$_\odot $, $\langle (B-V)\rangle =0.94$], that is, 
a mass--luminosity ratio 4.4 times larger. Assuming a variation of the luminosity
linearly dependent on the time, the variation of this ratio would be by a
factor 5.8 between z=0.1 (the average redshift of SDSS galaxies, which
are the reference of the size calibration in Shen {\it et al.}\cite{She03}) and $z=3.2$.
This is an acceptable estimation for elliptical galaxies. 
Kauffmann {\it et al.}\cite{Kau03}(Fig. 13) showed that SDSS late-type galaxies have
a mass--luminosity ratio around 2 times lower than early-type ones,
and Rudnick {\it et al.}\cite{Rud06}(Fig. 9) showed that blue galaxies at $z\approx 3$
also have mass--luminosity ratios around 0.2-0.3, so
we also keep this number of $\approx 6$ for disk galaxies.
This factor of 6 might be affected by important errors (see \S \ref{.statothers}), 
and could be much lower, though not much larger.

Given that we are comparing galaxies with the same luminosity,
the galaxies at $z=3.2$ would have 6 times lower stellar mass than at $z=0.1$.
Hence, from Eqs. (17)-(18) of Shen {\it et al.}\cite{She03}, we find that galaxies at
$z=3.2$ should be 2.7 times smaller than at $z=0.1$ if they are elliptical,
or 2.0 if they are disk galaxies.
These factors are much lower than the measured values of 6--10 and 4.5--6
respectively. A factor in size 2--4 for elliptical or 2--3 for disk galaxies, 
including this luminosity evolution correction
remains (in rough agreement with the results by Trujillo {\it et al.}\cite{Tru06}). 
Therefore, the argument of ``younger population in higher redshift'' 
does not serve to justify the present data. In order to explain
the observed size increase in terms only of luminosity evolution of
the stellar population, we would need to set $M_*/L_V$
with respect to the SDSS galaxies a factor 25--60 for elliptical
galaxies, or 50--100 for disk galaxies, too much!

\subsection{Mergers}
\label{.mergers}

If the luminosity density is not due
to an increase of the luminosity of each star, it must be due to
a redistribution of the mass density to make it more compact.
Why? Explanations in terms of mergers proliferate in
the literature (see discussion by Refs. \cite{Tru06,Tru07} and references
therein). Refs. \cite{Tru06,Tru07}
support the merger scenario calculating the size vs. mass ratio and 
observing how it also decreases with redshift\footnote{In 
Refs. \cite{Tru06,Rud06}, the masses were estimated 
from the colors of the galaxies assuming solar metallicity
for the determination of the mass--light ratio and a
Salpeter IMF model. Trujillo {\it et al.}\cite{Tru07} 
multiply the masses by a factor 0.5 to correct for
the difference of Kroupa and Salpeter IMF, which is only
a very rough approximation. The masses of Trujillo{\it et al.}\cite{Tru07} are more
accurately calculated, with an uncertainty of a factor  two.
All these assumptions introduce significant errors into the calculation
of the mass--light ratio. Therefore, we must take these results with care.},
but this calculation also depends on the cosmological model used
and can give different results for different cosmologies.
Furthermore, Khochfar {\it et al.}\cite{Kho06} point out that the presence
of higher amounts of cold gas at high redshift mergers of ellipticals
also produces size evolution.

Trujillo {\it et al.}\cite{Tru07} cites the paper of Boylan-Kolchin {\it et 
al.}\cite{Boy06} as a possible powerful mechanism to increase the radius 
in massive elliptical galaxies. 
Each merger would follow a law $R\propto M_*^{-\alpha }$ \cite{Boy06} 
with $\alpha =1.0$ for a pericenter 
of 15 kpc and lower for higher pericenter distances or higher
for lower pericenter distance. So when
two galaxies of the same luminosity approach each other reaching
a minimum distance of 15 kpc and merge, 
the total radius will be $\approx 2.00$ times the radius of each individual 
galaxy, while Shen {\it et al.}\cite{She03} for local galaxies 
give $\alpha =0.56$: a factor $\approx 1.47$ in radius on average
when we double the mass.
This means that each major merger (fusion of galaxies of the same luminosity)
gives an extra factor of 1.36 for ellipticals
in angular size with respect to eq. (\ref{shen2}) relationship.
We would need an average of 2.3--3.6 
($=\frac{ln(2-3)}{ln(1.36)}$, i.e. $1.36^{2.3-3.6}=2-3$)
major mergers along the life of each elliptical galaxy to justify 
a factor 2--3 in radius.

The observed rate of mergers is indeed much lower than the necessary rate
to justify these numbers. Lin {\it et al.}\cite{Lin04}
with  statistics of close galaxy pairs ($r<20 h^{-1}$kpc)
up to z=1.2 show that only $\sim 9$\% of the luminous galaxies 
($-21<M_B<-19$) would have 
a major merger during their lives since $z=1.2$. 
De Propis {\it et al.}\cite{DeP07} get similar merger ratios for local galaxies
($z<0.12$) than Lin {\it et al.}\cite{Lin04} for $0.5<z<1$.
The number of cumulative mergers increases up to 22\% 
\cite{Lin08} if we allow larger pair separation ($r<30 h^{-1}$kpc) 
and up to 54\% \cite{Lin08} if we allow
a broader definition of major merger including pairs of galaxies
with mass ratio up to four.
Ryan {\it et al.}\cite{Rya08} calculated a number of 42\% galaxies 
undergoing some merger up to $z=1$ for 
more luminous galaxies ($M_B>-20.5$) [and 62\% for all $z$],
also for mass ratios up to four, and $r<20 h^{-1}$ kpc.
In our case, comparison with major mergers of equal mass
galaxies and with average pericenter around 15 kpc should be made,
so a number of 10--20\% (up to $z=3.2$ is $\sim 50$\% larger than up to
$z=1$) of mergers would be the amount
to compare with the 2--4 mergers (200--400\%) we need for the size evolution.
Or we may account for mergers with mass ratios up to 4 and pericenter
radii larger than 15 kpc on average, although these
would not produce an increase of an extra factor of 1.36 for ellipticals
in angular size with respect to the Eq. (\ref{shen2}) relationship but 
lower. Therefore, we would have to compare the number of $\sim 50$\%
with the large number of mergers of this kind to produce the
observed evolution in radius ($\sim 10$ mergers per galaxy;
probability of merging $\sim 1000$\%).
 
Moreover, I suspect that the number of mergers is
overestimated. First because not all galaxy pairs
become mergers. Second, because of the contamination of 
interlopers. There is an uncertainty of radial distance due
to proper motion of the galaxies apart from the Hubble flow,
and many of the identified pairs of galaxies in the projected
sky are not real pairs in 3D space. There may be a chance superposition
in the line of sight of galaxies. The situation is worst
for Ryan {\it et al.}\cite{Rya08}, who use 
spectrophotometric redshifts; and interlopers do not introduce
a statistical error, as they say, but a systematic one.
However, the order of magnitude of the pairs with or 
without interlopers should be more or less the same,
the interlopers being  at $z<3$ lower than $\sim 30$\% \cite{Ber06}. 
Also, De Ravel {\it et al.}\cite{DeR08}, for instance, using 
spectroscopically confirmed pairs, get similar merger ratios.
Third, because timescales of mergers increase slightly with redshift 
and are longer than assumed in most observational studies\cite{Kit08}.

About the CAS method of estimation of merger ratios based
on the identification of major mergers with highly asymmetric
galaxies\cite{Con00,Con08} two things may be commented:
1) The authors consider that all asymmetries in principle associated
with starbursts are due to major mergers, but the mechanisms
which trigger important amounts of star formation might be
different from major (ratio 1:1) mergers; in particular, 
minor mergers may also trigger asymmetric star formation.
Strongly disturbed systems, indicative of recent strong
interactions and mergers, account for only a small fraction 
of the total star formation rate density\cite{Jog08}. 
2) Interlopers, galaxies and stars
with very different redshifts projected as background 
or foreground objects in the line of sight 
produce an important amount of apparent distortion in the galaxies,
especially at high redshift, where Conselice {\it et al.}\cite{Con08} claim
that a high fraction of galaxies are major mergers. We must bear in mind
that Hubble images may detect very faint galaxies and 
there are more than 5 million galaxies per square
degree with $m_z<30$ (\cite{Ell07}; extrapolated from
the counts up to magnitude 28),
one galaxy in each square of $1.6''\times 1.6''$ on average,
which, mixed with the main galaxy ($m_z<27$ in Conselice {\it et al.}\cite{Con08}),
produce apparently distorted galaxies.
Also, De Propis {\it et al.}\cite{DeP07} checked that the contamination is 
happening in low redshift galaxies with foreground stars.
Moreover, we see the galaxies once they are
``presumably'' merged but we do not see enough galaxy pairs at $1<z<2$ 
when both galaxies of the merger are separated (e.g., according to
the already overestimated ratio by Ryan {\it et al.}\cite{Rya08}). This 
indicates at least that the asymmetries are not produced by major
mergers but by minor mergers (ratio of masses larger than 4) or other
effects.

Stellar population analyses do not even agree with these merger rates. 
Mergers or captures of smaller galaxies can occasionally occur, 
but hierarchical-scheme  
subunits fusing together and made of gas and stars is not the dominant one by which 
massive elliptical galaxies are made, at least for $z<2$ 
\cite{Chi02}. Most massive elliptical galaxies
have a passive evolution since their creation according to stellar population
analyses\cite{Chi02}. Mergers are beautiful, spectacular
events, but not the dominant mechanism by which elliptical galaxies
are assembled.
Most early-type galaxies with a velocity dispersion exceeding 200 km/s 
formed more than 90\% of their current stellar mass at redshift $z>2.5$
\cite{Jim07}.
Elliptical galaxies formed in a process similar to monolithic 
collapse, even though their structural and dynamical properties are 
compatible with a small number of dry mergers\cite{Cio07},
far from the number of mergers necessary in our case.
Dry mergers do not decrease the galaxy stellar-mass surface
density enough to explain the observed size evolution\cite{Nip09}, and the high density
of the high-z elliptical galaxies does not allow them to evolve into present-day 
elliptical galaxies\cite{Nip09}.

Mergers are searched for in the Local Group galaxies too.
In the case of the Milky Way, some authors try
to find evidence of major fusion events  of big galaxies, 
but up to now we do not see evidence in favour but against such scenarios\cite{Ham07}. 
There are minor mergers, of course, and
absorption of small clouds of the intergalactic medium, but
the presence of intermediate mass galaxies
at short distances from the center, at present or
in the past, have yet to be identified. There are certain attempts
to find something, for instance the recent discovery of a galaxy 
with a relatively large diameter at only 13 kpc from the 
Galactic center called Canis Major, but
that discovery resulted in a fiasco\cite{Lop07}.
In any case, two to four major mergers on average per galaxy is too much.

Another argument against the merger scenario of the hierarchical
CDM cosmology is that galaxy formation is controlled
by a single parameter\cite{Dis08}. 
One would expect in the merger scenario that the properties of individual
galaxies be determined by a number of factors related to the star
formation history, merger history (masses, spins and gas content of
the individual merging galaxies), etc., but that is not so; all the
different parameters of the galaxies are correlated\cite{Dis08} and
there is only one single independent parameter based on their mass.

On the other hand, major mergers of disk galaxies of comparable
mass should give place to elliptical galaxies, so it is not easy to understand
in this scenario how the radius of disk galaxies grows.
Conselice {\it et al.}\cite{Con05} show in fact that there is little to no evolution
for disk galaxies at $z<1.2$, for the K-band, in the stellar-mass 
Tully--Fisher relation, and in the ratio of stellar/total mass. 
Ferguson {\it et al.}\cite{Fer01} also
find that accretion flows play only a minor role in determining the evolution 
of the disk scalelength. In models in which the main infall phase precedes 
the onset of star formation and viscous evolution, they find the exponential 
scalelength to be rather invariant with time. On the other hand, 
models in which star formation/viscous evolution and infall occur 
concurrently result in a smoothly increasing scalelength with time, 
reflecting the mean angular momentum of material which has fallen in at 
any given epoch.

Furthermore, selection effects go apparently in the opposite 
direction of observing pre-merger galaxies at high redshift. 
At very high redshift, we are observing only galaxies with
stellar masses over $10^{11}$ M$_\odot $ and some of them 
over $10^{12}$ M$_\odot $ \cite{Tru06}.
And at low redshifts there are galaxies of all masses but the average
stellar mass is much lower than that.
Thinking that very massive galaxies
at high $z$ are building blocks of even more massive low $z$ galaxies 
is counterintuitive. After 2--4 mergers of equal mass
galaxies in average, the galaxies should be 
4--16 times more massive than the original
building blocks at high redshift. We should be observing some galaxies at
low redshift with stellar masses of $\sim 10^{13}$ M$_\odot $. 
With a mass-luminosity ratio of $M_*/L_V=6$ 
(for a very old population \cite{Vaz96}; 
if it were younger than 12 Gyr it would be lower so the
luminosity would be even higher),
this would mean galaxies with $L_V=2\times 10^{12}$ L$_{\odot ,V}$,
or absolute magnitude $M_V=-26$. Even cD galaxies in the centers of 
the clusters are not as bright as that. Where are these galaxies, then?

Thinking that low- to intermediate-mass galaxies are 
the final stage of major merger processes is a reasonable possibility,
since we cannot see their building blocks at high redshift 
(they are very faint). However, for only high luminous
galaxies I get more or less the same shrinking factors at high-z 
with respect to low-z (see Fig. \ref{Fig:angsizelum}).
It is not a question of some merging which affects low
luminosity galaxies more. I could even concentrate our analysis on galaxies with 
$L_V>1.02\times 10^{11}$ L$_{\odot ,V}$ and, in spite of the poorer statistic,
a very strong size evolution between galaxies
at $z=0.5-1$ and $z>2$ can be appreciated. Refs. \cite{Tru07,Bui08}
even get higher evolution for higher stellar masses, but 
this has been criticized\cite{Man09}.

Another element that is not consistent with these hierarchical merging scenarios 
is that superdense massive galaxies should be common in the early 
universe ($z>1.5$), and a non-negligible fraction (1-10\%) of them 
should have survived since that epoch without any merging process
retaining their compactness and presenting old stellar populations in 
the present universe. However, Trujillo {\it et al.}\cite{Tru09} find only a tiny 
fraction of galaxies ($\sim 0.03$\%) of these superdense massive galaxies 
in the local Universe ($z<0.2$) and they are relatively young ($\sim 2$ Gyr)
and metal-rich ($[Z/H]\sim 0.2$). Clearly a case of how some authors
(Trujillo {\it et al.}\cite{Tru09}) 
try to find proofs in favour of the hierarchical merging
scenario, and when they find that the observations point out exactly 
the opposite thing of what is expected, instead of claiming that the 
hierarchical merging scenario is wrong, they try to deviate attention by
giving less importance to the observed facts for the validation of
the standard theory.
 
\subsection{Quasar feedback}

Fan {\it et al.}\cite{Fan08} realized that the evolution+merger luminosity 
solution is not enough to explain the strong size evolution and they
claimed, ``no convincing mechanism able to account for such
size evolution has been proposed so far''. Nonetheless, 
they have proposed a new mechanism to explain this extraordinary 
size evolution. Fan {\it et al.}\cite{Fan08} propose that in elliptical galaxies it 
is directly related to a quasar feedback:
part of the energy released by the QSO would be spent to produce
outflows of huge amounts of cold gas expelled from the central regions, 
a rapid (few tens of Myr) mass loss which induces an expansion of 
the stellar distribution. 

Although this mechanism might  explain 
some small part of the expansion of elliptical galaxies, 
there are some aspects in this hypothesis which do not fit  
with the observed facts well, at least while we do not clarify some points. 
First, we do not see such supermassive outflows which are necessary to
maintain the Fan {\it et al.}\cite{Fan08} idea, although since
their life is very short only a few high redshift 
QSOs would show it. Second, elliptical galaxies expend their gas to
produce stellar formation which gives rise to their stellar mass.
If a QSO swept the gas away, no young stellar populations
would be observed, but there is now compelling evidence for a 
significant post-starburst population in many luminous AGN\cite{Ho05}.
There is also detection of large amounts of warm, extended 
molecular gas indicating that QSOs have vigorous star formation\cite{Wal07}, 
and that the gas is not being expelled.
The blue color of host galaxies, $(B-V)_{rest}\sim 0$ \cite{Sch08},
indicates a young population too. 
Third, if QSOs  produced such massive outflows ejecting
the gas of the galaxies, this would also apply to disk galaxies.
Around 40\% of host galaxies in QSOs are disk galaxies\cite{Guy06}, 
and it is clear that disk galaxies still have  gas
and active star formation in their disks.
Fourth, a continuous increase in the average size of ellipticals,
as shown in Fig. \ref{Fig:angsizetype}/left, would require the continuous
expulsion of gas, but most massive elliptical galaxies
have had a passive evolution since their creation at $z>2$ 
according to stellar population analyses\cite{Chi02}, which
indicate that the gas was already drained in them at $z>2$.
Fifth, for an average increase of a factor 3 in size  in the elliptical
galaxies, we would need to assume that most elliptical galaxies have
hosted very luminous QSOs during their lives. With a minimum QSO lifetime of 
around 40 Myr, as required for the massive outflow mechanism of
Fan {\it et al.}\cite{Fan08}, the number of very luminous QSOs should be 
around 1/300 of the number of elliptical galaxies. 
This number is too large, given that the density of QSOs
with $L_{bol}>\sim 10^{48}$ erg/s (the necessary luminosity,  
5\% of which is spent to produce outflows larger than 1000 M$_\odot /yr$, as posited by
Fan {\it et al.}\cite{Fan08}; $L\sim \frac{M\dot{M}}{0.05R}$ with
$M>2\times 10^{10}$ M$_\odot $ and $R=\frac{1}{3}R_{SDSS}$) 
is $\sim 10^{-8}$ Mpc$^{-3}$ \cite{Hop07}.

Nonetheless, I would not dare to say that Fan {\it et al.}'s\cite{Fan08} hypothesis 
is incorrect. I think it is an elegant and interesting solution to the problem,
and it must be considered as a serious possibility, provided it is
able to solve their caveats, and give some support to the scenario
with some observations directly interpretable as massive outflows.

\subsection{Rotation or dispersion velocity analysis}

If this hypothesis of lower radius at high redshift
for a given mass were true, due either  to mergers or quasar feedback,
we would expect a significant increase in the
rotation speed or dispersion velocity 
of those galaxies at high redshift with respect to the
local ones.  For a constant mass within a radius $R$, one would roughly
expect rotation speed in a galaxy  
$v_{rot}\propto R^{-1/2}$, and something similar for the 
dispersion velocities in ellipticals. 
Let us analyze whether this change of velocity is taking place.

\begin{figure*}
\vspace{1cm}
{\par\centering \resizebox*{5cm}{5cm}{\includegraphics{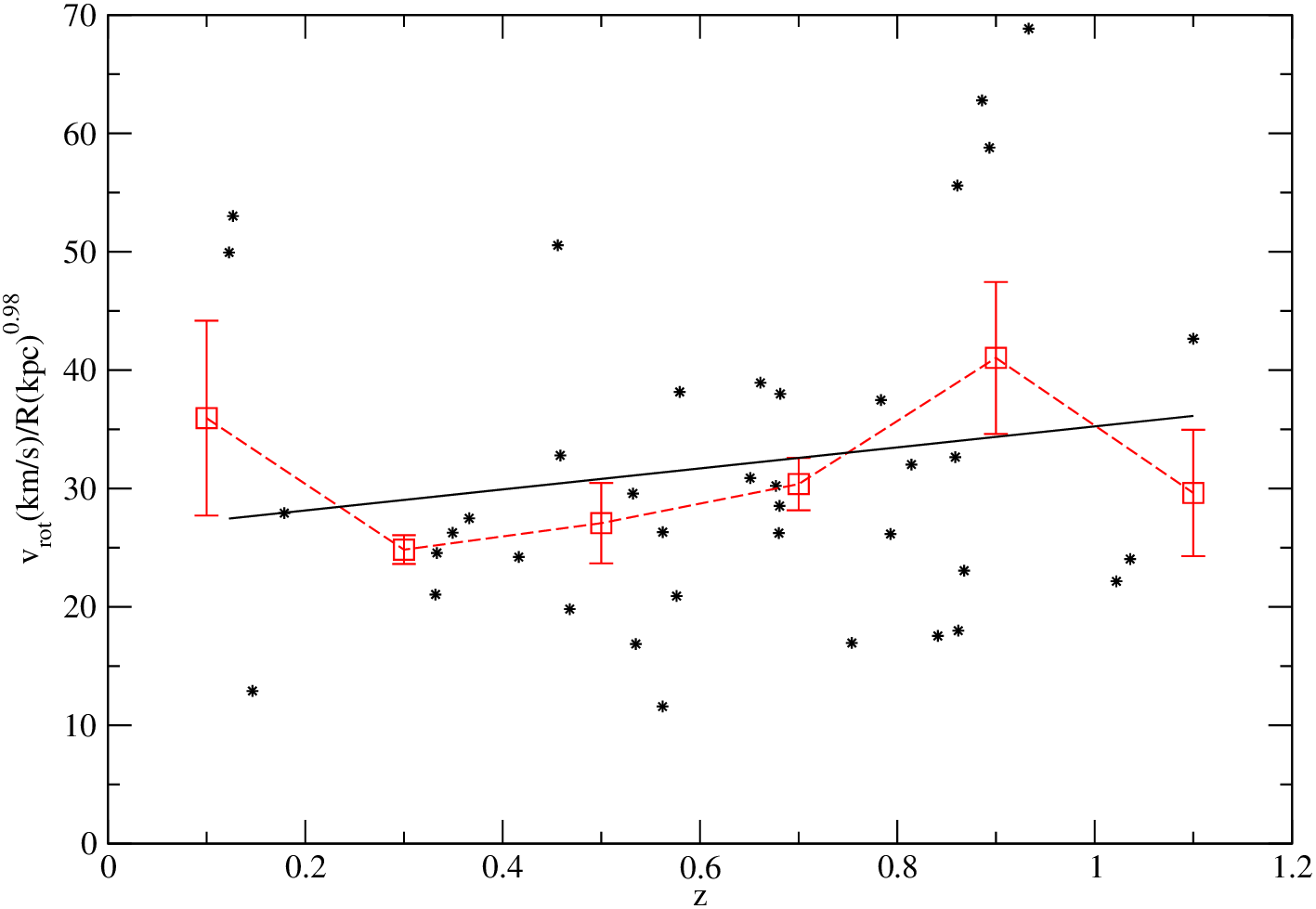}}
\hspace{1cm}\resizebox*{5cm}{5cm}{\includegraphics{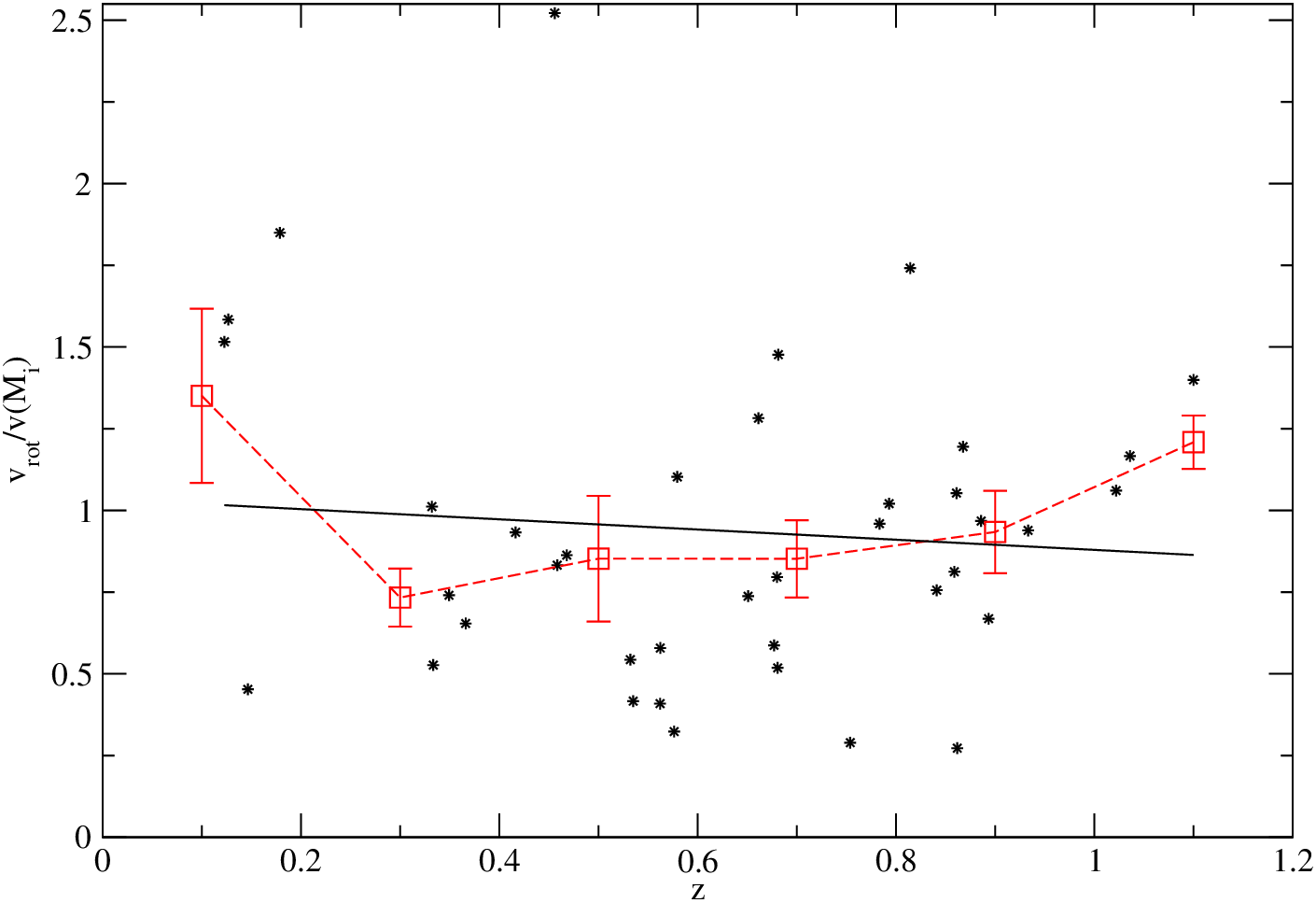}}\par}
\caption{}
Plot of $v_{rot}/R^{0.98}$ and $v_{rot}/v(M_i)$ for the 39
galaxies of the sample from Marinoni {\it et al.}\cite{Mar08b}. Squares with error
bars are the average of the data (asterisks) with steps of $\delta z=0.2$.
The solid line is the best linear fit, given by eqs. (\ref{vmar1}) 
and (\protect{\ref{vmar2}}).
\label{Fig:marinoni}
\end{figure*}

An analysis of Marinoni {\it et al.}'s \cite{Mar08b}
data with galaxies with $z<1.2$ does not show (see Fig. \ref{Fig:marinoni}) 
a significant change in the rotation velocity--size relationship or 
the rotation velocity--absolute 
magnitude relationship. According to the Saintonge {\it et al.}\cite{Sai08} analysis for
low redshift galaxies, the rotation velocity is proportional 
to $R^{0.98\pm 0.01}$, and  is also related to the absolute magnitude in the
I-band by their eq. (3), so we could derive an average expected
velocity from the luminosity of the galaxy, $v(M_i)$. In Fig. 
\ref{Fig:marinoni}, I see that these relationships remain more or less
constant: the variations in $v_{rot}/R^{0.98}$ and $v_{rot}/v(M_i)$ 
are compatible within 1-$\sigma $ to be null. Particularly,
the best linear fits give:
\begin{equation}
\frac{v_{rot}}{R^{0.98}}=(26.4\pm 5.9)+(8.9\pm 8.8)z
\label{vmar1}
,\end{equation}
\begin{equation}
\frac{v_{rot}}{v(M_i)}=(1.04\pm 0.20)-(0.09\pm 0.16)z
\label{vmar2}
.\end{equation}
Within the error bars, I cannot exclude an increase
in these ratios with redshift compatible with the hypothesis
of radius decrease. An interesting test would be to measure the
rotational velocity in some of the very compact galaxies with 
redshift 3. Although getting a spectrum of these faint galaxies 
is technically difficult, this would give a proof of whether either they
are really so compact or the cosmological parameters are wrong.

The critical assumption of a variable effective
radius is also counter-argued by the proofs in favour of a constant
radius\cite{van98} showing that high redshift first-rank
elliptical galaxies, with similar absolute magnitudes,
have the same velocity dispersions as low redshift 
first-rank elliptical galaxies. 
Van Dokkum {\it et al.} \cite{van98} point out, however, that
these galaxies are predicted by galaxy formation models to be those
whose formation finished at very high redshift and so it would not be
surprising that these galaxies had the same radius as at low
redshift, specifically because they are in clusters where mergers
may not be likely.

A surprising new result also points in this direction\cite{Cen09}: 
the velocity dispersion of giant elliptical galaxies 
with average redshift $z\approx 1.7$ from Cimatti {\it et al.}'s \cite{Cim08} sample is 
similar or very slightly larger than the dispersion  for the same kind
of galaxies with the same stellar mass in the local Universe at
 240 km/s, while at $z=0$ it is around 180 km/s. Since
Cimatti {\it et al.}\cite{Cim08} galaxies at $z\approx 1.7$
are more compact than the average size with that luminosity at that redshift, 
it is normal to have a slightly higher velocity dispersion. 
It should be much higher (over 400 Km/s) at high redshift since a 
much lower radius is attributed to them \cite{Cim08}, but it is not.
This result points directly to the conclusion that the galaxies have
not strongly changed  their radii. Other alternative ad hoc  ideas in terms of 
a conspiracy of effects in which the dark matter ratio has increased
the amount necessary to compensate for the radius increase\cite{Cen09} 
sound like a queer coincidence, 
and have no clear basis in terms of galaxy formation scenarios.

\subsection{Discussion on expansion+evolution models}

All these considerations may make us think that the concordance model
cannot explain the present data. The other expanding models present 
similar problems: a factor in average size evolution up to $z=3.2$ equal
to 5.8 for Einstein--de Sitter, 5.6 for Friedmann 
with $\Omega _m=0.3$, and 5.9 for the QSSC model.
Phenomenologically, it is possible
to fit the data to any of the expanding models with appropriate evolution of
galaxies. In practice, this evolution for a constant 
luminosity galaxy should be so strong (up to 
a factor 200 in average in the luminosity density up to $z\approx 3.2$;
systematic errors may change up to a factor 2 this number, but
this does not change the situation) 
that the explanations for it seem unrealistic.

The situation for the expanding models becomes even more dramatic
if we go to higher redshift. At redshift 6, the linear
size of the galaxies assuming a concordance model is even lower, 
approximately a factor two lower than at $z=3.2$ (\cite{Bou04} Fig. 4). 
If we were going to consider a reduction in size by
a factor 12, all the arguments given in this section would become
even stronger.

\section{Analysis of the static universe cases}

In the first two  static Universe cases, 
there is an excess of size ($\sim 20-30$\%)
for most redshifts.
A possible interpretation of this discrepancy is that there 
may be a systematic error in the calculation of the ratio between measured
and predicted average $\theta _*$. In \S \ref{.select}, I have
discussed the possible systematic errors and I concluded that 
they should be lower than $\sim 30-40$\%.
These systematic errors might be enough to justify the departures
of the data with respect to the prediction in Fig. \ref{Fig:angsize} 
for these static models. 

The third case of plasma redshift is much more discrepant
and can only fit the data with a significant evolution of galaxies, although
less strong than in the expanding models: a factor 3 instead of a factor 6 in the
concordance model for a given luminosity from $z=0$ to $z=3.2$.

\subsection{Including extinction}
\label{.extinc}

Apart from the systematic error considerations,
another solution to make  the two first static cosmological models
compatible with our data would be related to  extinction
rather than the evolution.
Extinction would make the galaxies  look fainter, 
which means that, through Eqs. (\ref{thetaequiv}) and
(\ref{shen2}), when the corrections
of extinctions are made, the luminosity is larger and
$\theta _*$ is smaller than their values without
corrections. Let us check this hypothesis with a rough calculation.
Instead of Eq. (\ref{lrest}), the inclusion of the 
IGM extinction with absorption
coefficient $\kappa $ (area per unit mass) will give the following
relationship

\begin{equation}
L_{V,rest}=4\pi F_{V,rest}d_L^2e^{\rho _{dust}\int _0^{d_A}dr\ \kappa
[\lambda _V\frac{1+z(d_A)}{1+z(r)}]}
.\end{equation}
I have assumed a constant dust density $\rho _{dust}$ along the line
of sight, which is an appropriate approximation for a homogeneous
Universe without expansion and moderate amounts of dust ejection by the galaxy. If
we considered an expanding Universe with a strong dust emission rate
by the galaxies, we should include $\rho _{dust}(r)$ within the integral,
but it is not the case here.
The absorption coefficient can approximately be described with a
wavelength dependence:
\begin{equation}
\kappa (\lambda )=\kappa (\lambda _V)\left(\frac{\lambda }{\lambda _V}
\right)^{-\alpha }
.\end{equation}
Hence, and using Eqs. (\ref{angdistst}) and (\ref{angdisttl}),
\begin{equation}
L_{V,rest}=4\pi F_{V,rest}d_L^2
e^{\frac{c\ a_V}
{H_0(\alpha +m)}[(1+z)^m-(1+z)^{-\alpha}]}
,\end{equation}
with $m=1$ for the cosmology with linear Hubble law, and
$m=0$ for the tired light case. $a_V\equiv \kappa (\lambda _V)\rho _{dust}$
is the absorption in V per unit length, which means
there are $1.086a_V$ magnitudes in V of extinction per unit length.

The value of the exponent
$\alpha $ is not totally independent of $\lambda $ but I take it 
approximately as constant. I adopt $\alpha =2$, as observed
in near-infrared bands in our Galaxy\cite{Nis06}.
For lower wavelengths (optical, ultraviolet) $\alpha $ would be lower.
Since most of the sources have the
wavelength equivalent to V-rest in the near-infrared, and
the extinction curve of dust in the intervening QSO absorbers 
resembles the SMC extinction curve\cite{Kha05}, 
this approximation is reasonable.

The absolute value of $a_V$ is not well known and neither do we
know whether it is significant or null. There are only
some constraints for the maximum value (e.g., \cite{Ino04},
although based on standard cosmology). 
The values of $a_V$ which give the best fit
to our data are: $a_V=1.6\times 10^{-4}$ Mpc$^{-1}$ for the linear
Hubble law case, and $a_V=3.4\times 10^{-4}$ Mpc$^{-1}$ for the
simple tired light case. Assuming
$\kappa (\lambda _V)\sim 10^5$ cm$^2$/gr \cite{Wic96},
the value for the dust density necessary to produce such an extinction
would be $\rho _{dust}\sim 6\times 10^{-34}$ g/cm$^3$,
and $\rho _{dust}\sim 1.2\times 10^{-33}$ g/cm$^3$ respectively, 
which is within the range of possible values. Vishwakarma\cite{Vis02} gives values of 
$\rho _{dust}=3-5\times 10^{-34}$ g/cm$^3$, but for the QSSC model.
Inoue \& Kamaya\cite{Ino04} allow values as high as $\rho _{dust}\sim 10^{-33}$ 
g/cm$^3$ for the high z IGM within the standard concordance cosmology. 
For comparison, the average baryonic density of
the Universe is (taking $\Omega _b=0.042$; \cite{Spe07})
$\rho _b=3.9\times 10^{-31}$ g/cm$^3$, so this would mean that
IGM dust constitutes  0.15 or 0.30\% of the total baryonic matter, reasonable
amounts.

Whether the extinction used in the models with extinction would be grey or would 
introduce some small reddening is not totally clear, but this is not
a question for the present analysis. I just note that reddening in the optical would
depend on the value of $\alpha $ in the visible at intermediate to high redshift
and the variability of $\kappa (\lambda )$ with respect to $\lambda $ in the UV.
The features of dust extinction in the UV are not easy to model in an IGM with
unknown composition.

\begin{figure*}
\vspace{1cm}
{\par\centering \resizebox*{5cm}{5cm}{\includegraphics{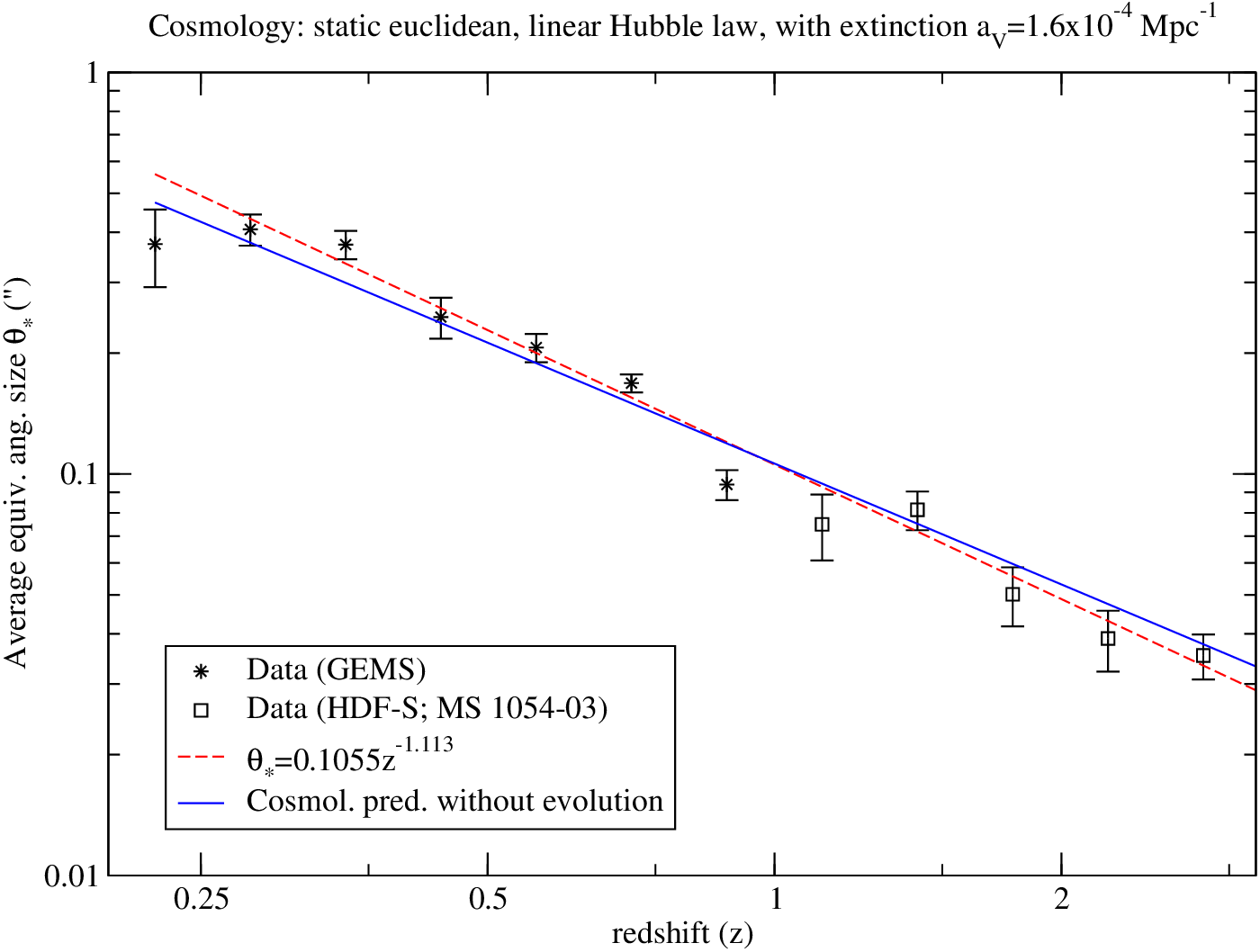}}
\hspace{1cm}\resizebox*{5cm}{5cm}{\includegraphics{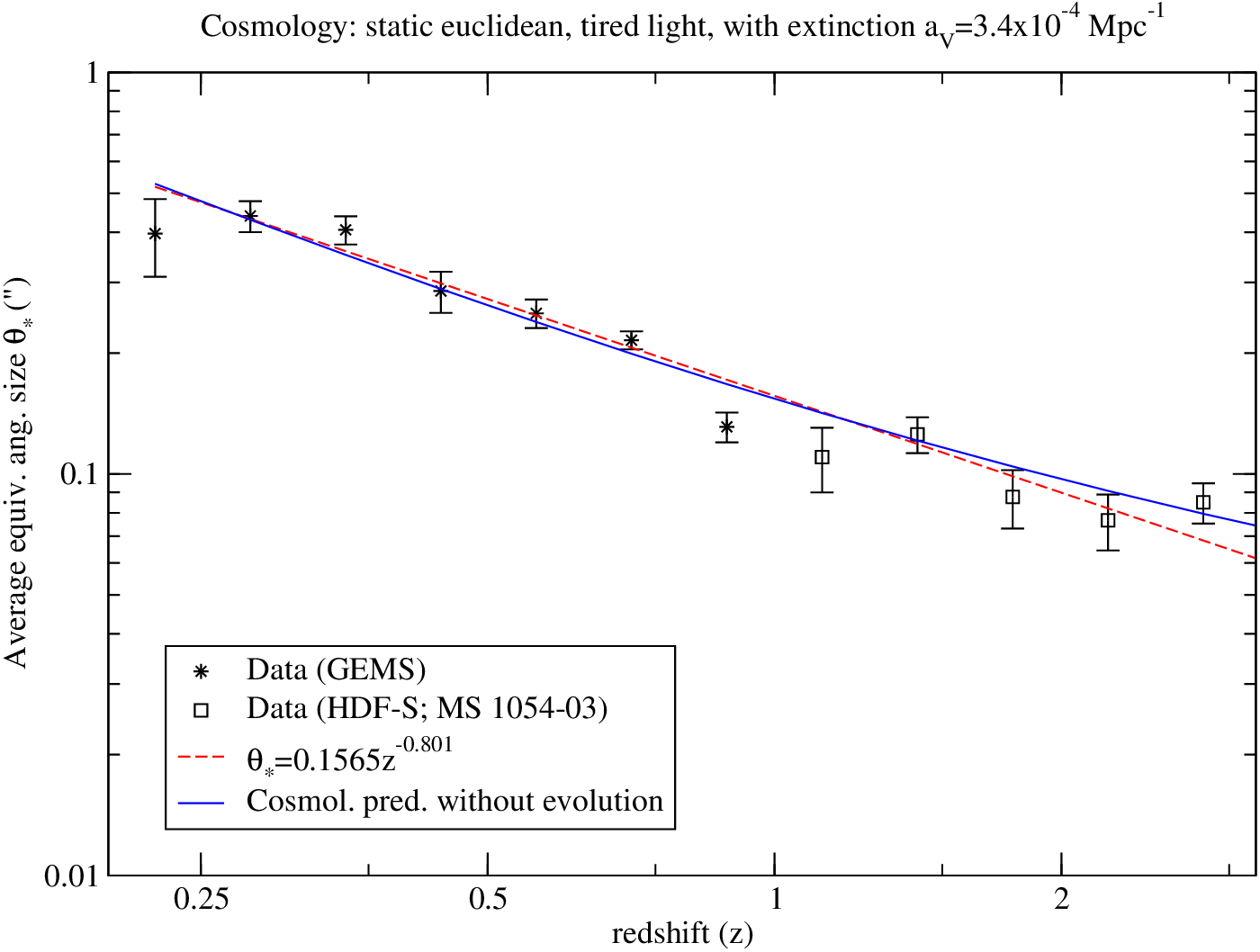}}\par}
\caption{Log--log plot of the average of $\log _{10} \theta _*$, where
$\theta _*$ is the equivalent angular size, vs. redshift ($z$). 
Bins of $\Delta \log _{10}(z)=0.10$.
The two plots are for the first two static cosmologies with the inclusion
of a constant IGM extinction which gives the best fit.}
\label{Fig:angsizeext}
\end{figure*}

With this simple correction for extinction, 
the results are significantly improved, 
as shown in Fig. \ref{Fig:angsizeext}. The tired light case
gives a better fit.

\subsection{Comparison with angular size test for ultra-compact radio
sources}
\label{.ultracom}

Compact radio sources have been used by several authors to carry out
the angular size test because these sources were thought to be 
free of evolutionary effects. However, the different results obtained
with these sources has raised the suspicion that they may not be
such good standard rods. Apparently, these rods are somewhat flexible.
For example, Kellermann\cite{Kel93} claimed that the angular size test 
for these sources fitted Einstein--de Sitter expectations  very well, 
when Einstein--de Sitter was the fashionable model. 
Jackson \& Dodgson\cite{Jac97} claimed the opposite: that it was not compatible with
Einstein--de Sitter, and that, given that $\Omega _m=0.2$,
the best fit for the cosmological constant term was 
$\Omega _\Lambda =-3.0$; flat cosmological models were excluded 
with $>70$\% C.L. Jackson\cite{Jac04}, 
in the era of the concordance model as the fashionable cosmology, 
again carried out the analysis of the same data used 
by Jackson \& Dodgson\cite{Jac97},
doing some new corrections due to selection effects and bias,
and they get the best fit for $\Omega _m=0.29$, $\Omega _\Lambda =0.37$,
compatible within 1-$\sigma $ with the concordance model.
With further data, Jackson \& Jannetta\cite{Jac06} get the
best fit for $\Omega _m = 0.25^{+0.04}_{-0.03}$, 
$\Omega _\Lambda = 0.97^{+0.09}_{-0.13}$ (68\% C.L.).
It seems that the general trend is to obtain the result
expected from fashionable cosmologies on the date in which the test is
carried out, and when incompatibilities appear, some selections
effects, biases, small evolution effects are sought to try to make the
results compatible. In my opinion, this is not a very
objective way to do science, but let us leave the discussion of the
methodology of cosmology aside.

One important selection effect is derived from the fact that
linear sizes depend on radio luminosities. Jackson\cite{Jac04} tries to
take this effect into account and suggests a method of correcting
it: binning the data into groups of 42 points in the redshift distribution
and taking as representative of each bin the mean of points between
11 to 17 within each one, counting from the smallest objects.
This is supposed to be done to compensate for the dependence 
$R_{rad}\propto L_{rad}^{-1/3}$ for a given redshift
and for the fact that the lowest luminosity points
cannot be observed at high redshift. In my opinion, this is
not the right way to correct the selection effect. Jackson\cite{Jac04}
is just doing a median which gives more weight to smaller objects,
but this median is shifted at high redshift by the lack of low 
luminosity objects. 

Nevertheless, my concern is not about Jackson's
method of correcting for the Malmquist bias, but about the
relationship between radius and luminosity. The relationship in Fig. 1
of Jackson\cite{Jac04} is applicable only to the concordance model
and it will be far different for static Universes. Particularly
in Figs. 1 and 2 of Jackson\cite{Jac04} we see that the linear size is
almost independent of redshift between $z=0.5$ and $z=4$: around
10 pc, with some scattering due in part to the range of luminosities for
each redshift. However, if I use eq. (\ref{angdistst}) of the static
Universe with linear Hubble law instead 
of eq. (\ref{concordance}) of the concordance model 
to calculate the angular size distances, I see that the
linear sizes at $z\approx 3.5$ should be a factor of 10 larger
($\sim 100$ pc instead of $\sim 10$ pc). Therefore, the linear
sizes would not be independent of redshift but highly dependent
on it. In such a case, Figs. 1 and 2 of Jackson\cite{Jac04} would be
transformed into a plot with a continuous increase in linear size with
radio luminosity for all ranges of luminosity for all redshifts: 
roughly $R_{rad}\propto L_{rad}^{1/2}$. 
Since at high redshift we cannot observe
objects with low $L_{rad}$, this means that we would be losing
objects of low $R_{rad}$ at high $z$, so 
the median of $\theta $ would be highly overestimated with respect
to low to intermediate redshift objects. 
This would explain why the
angular size test for ultra-compact radio sources does not
give a $z^{-1}$ dependence. Therefore, a static Universe is not
excluded by this test unless we demonstrate with data at low redshift
that the radii of the ultra-compact sources does not depend on their
luminosities.

Ultra-compact objects could be
used to carry out a right angular size test, but we cannot
directly compare objects of low luminosity at low redshift
with objects of high luminosity at high redshift
because we have no guarantee that the linear size is the
same in both cases (and it is not valid to assume a cosmological
model a priori to prove that it is good a standard rod, 
because the method should be independent of any cosmological
assumption if we want to derive from it which is the best cosmological
model). We should either i) compare objects of the same luminosity
(different for each cosmology), or ii) define a $\theta _*$ as in
the present paper in which we need to calibrate the size--luminosity 
relationship in the low-z Universe. This second option has the
caveat that we do not have very high luminosity compact radio sources
at low redshifts so we need to extrapolate the local radius--luminosity
relationship for high luminosities.

\subsection{Hubble diagram for the different cosmologies}
\label{.hubble}

\begin{figure*}
\vspace{1cm}
{\par\centering \resizebox*{5cm}{5cm}{\includegraphics{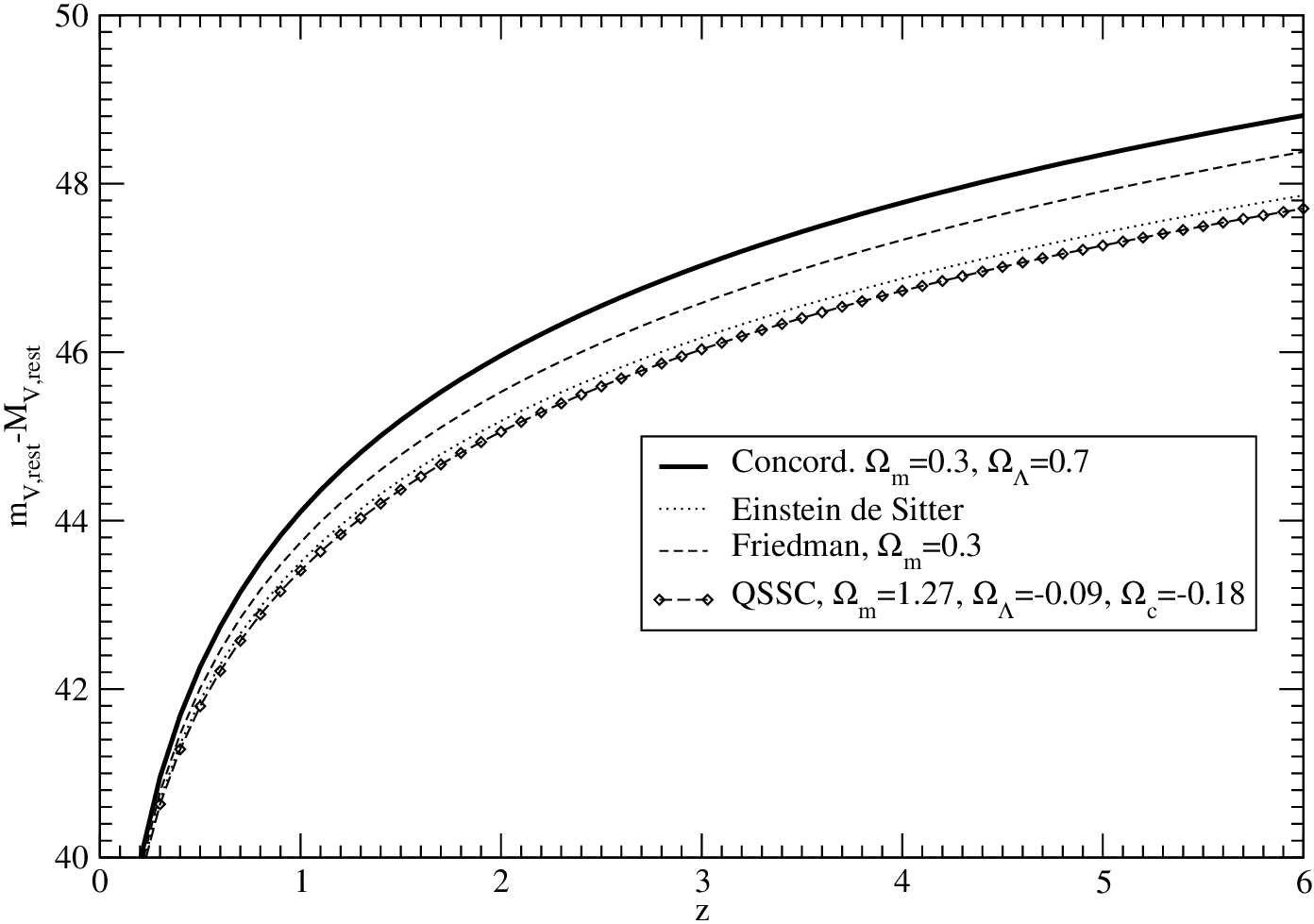}}
\hspace{1cm}\resizebox*{5cm}{5cm}{\includegraphics{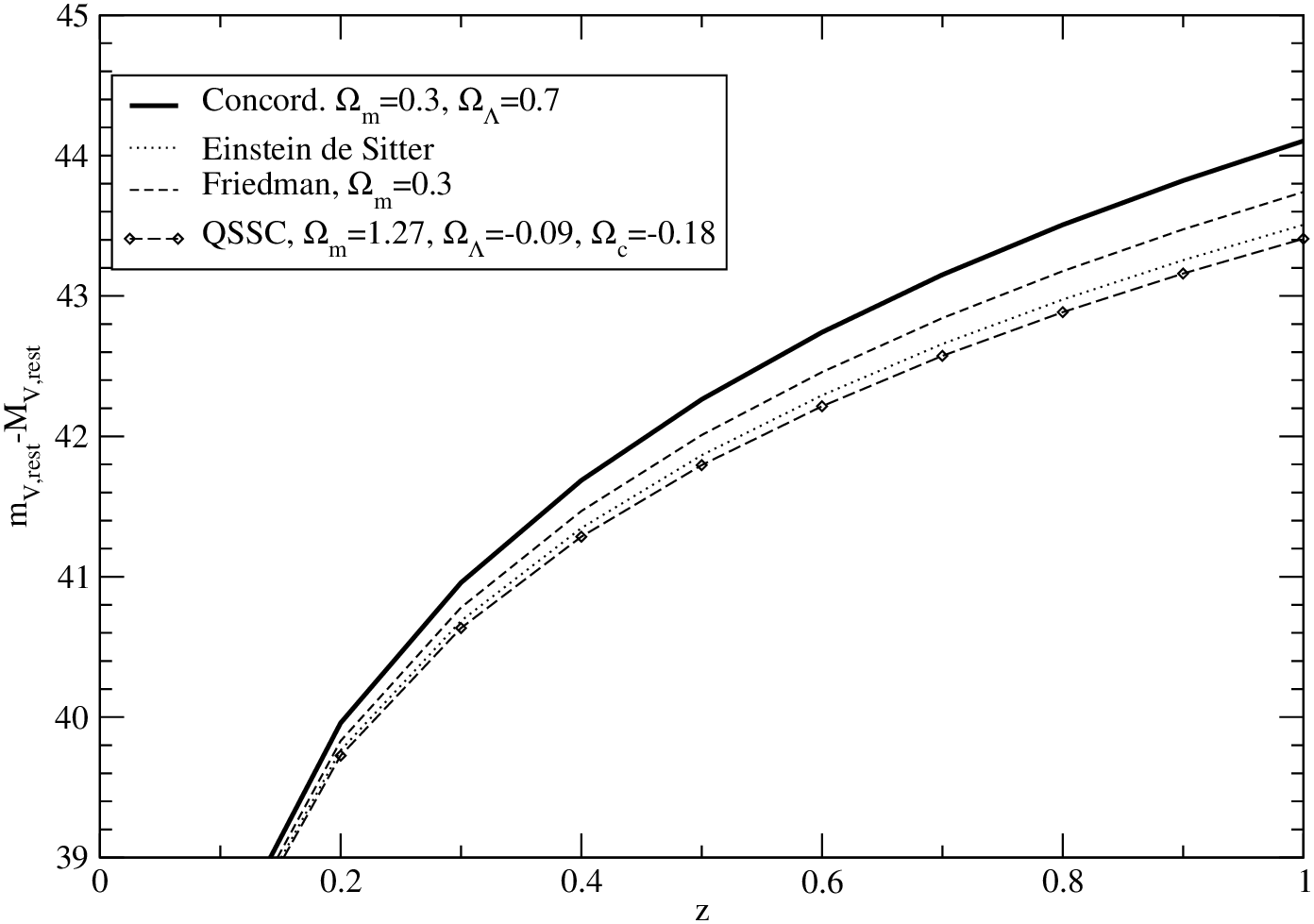}}\par}
\vspace{1cm}
{\par\centering \resizebox*{5cm}{5cm}{\includegraphics{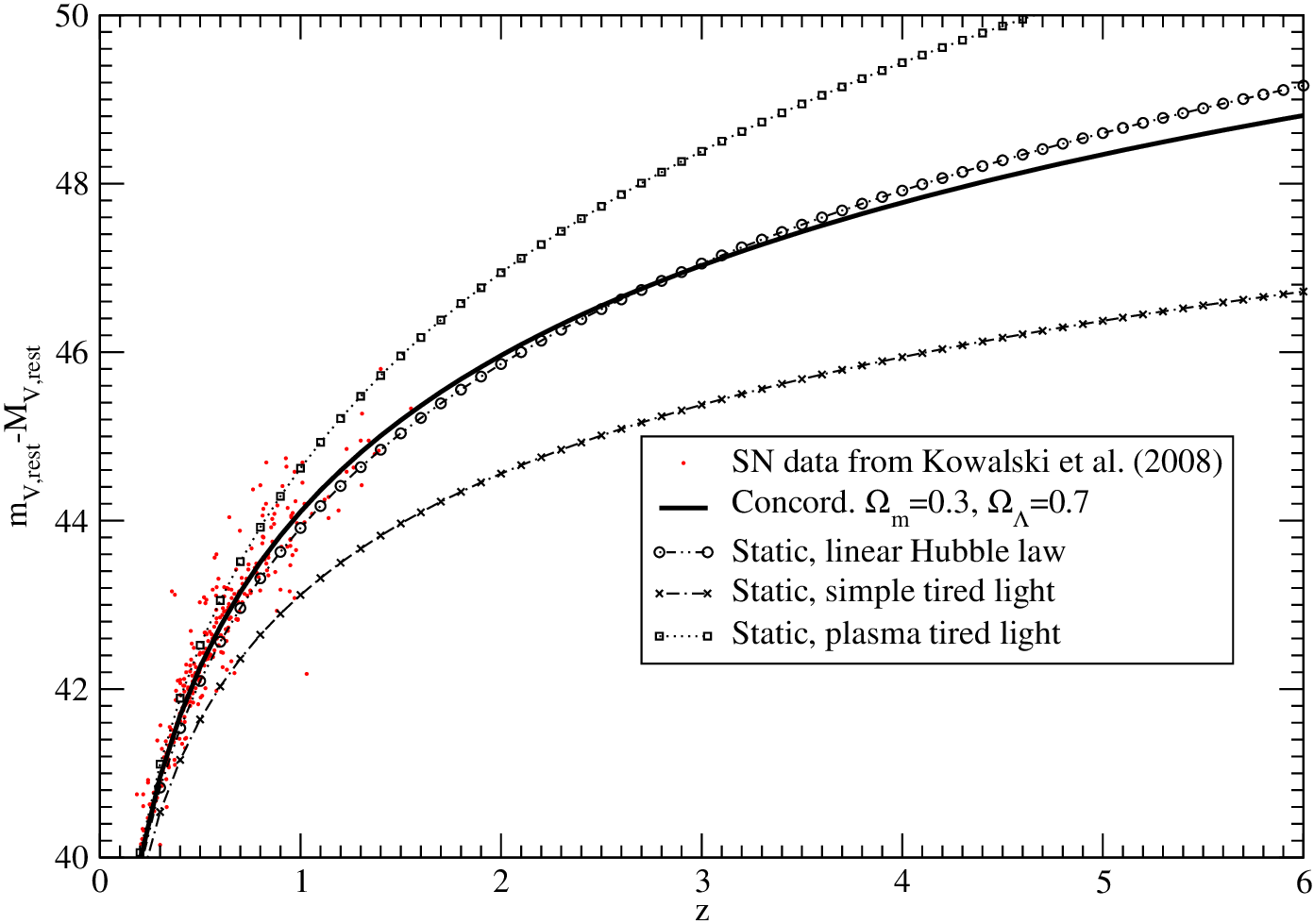}}
\hspace{1cm}\resizebox*{5cm}{5cm}{\includegraphics{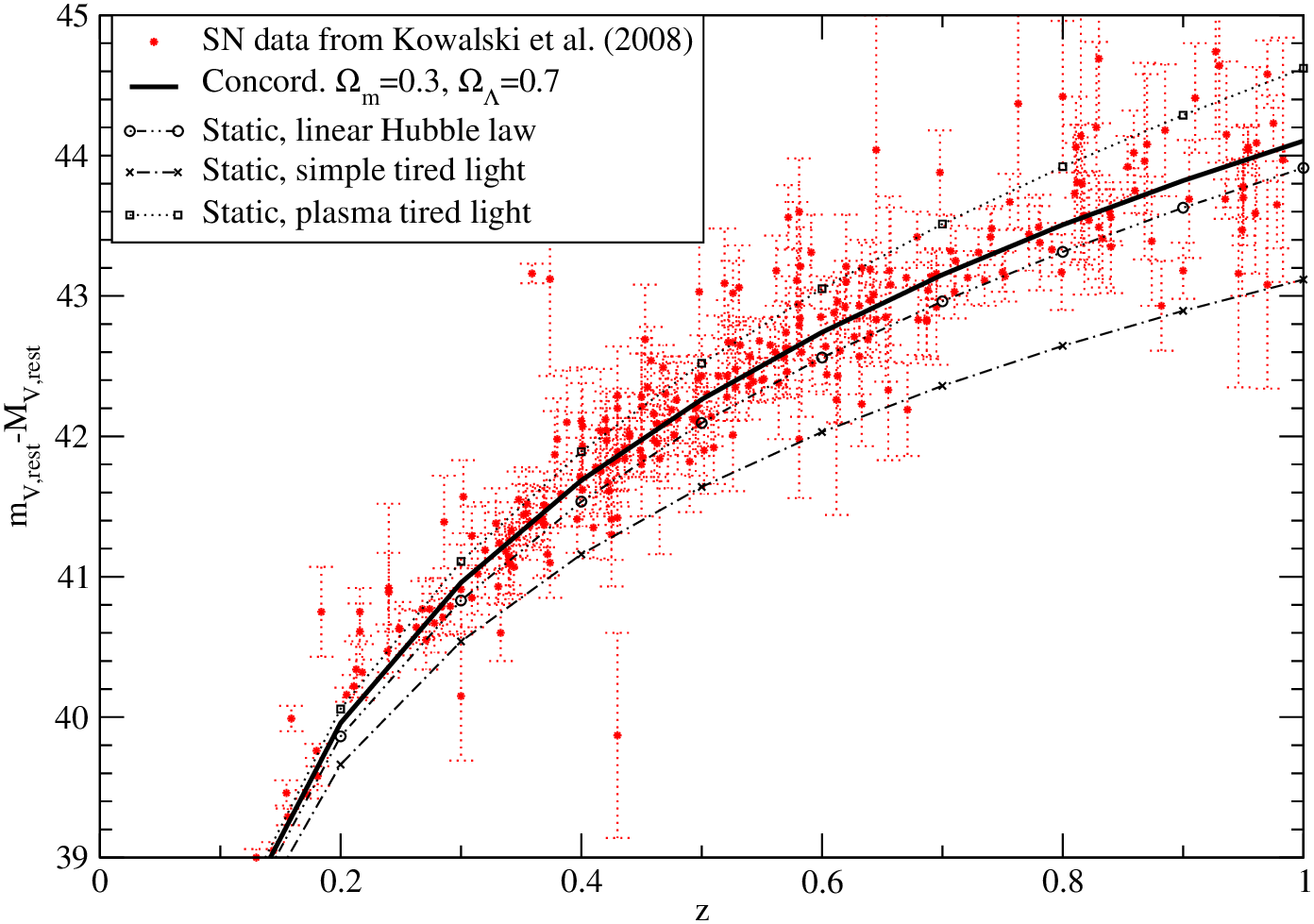}}\par}
\vspace{1cm}
\caption{}
Distance modulus as a function of redshift for different
cosmological models. Data of supernovae Ia compiled by Kowalski {\it et al.}\cite{Kow08} 
were added. Right plots are zooms of the left plots.
\label{Fig:hubble}
\end{figure*}

In Fig. \ref{Fig:hubble}, I show the different distance moduli
for the different cosmologies without extinction, together with some real
data of SNe Ia compiled by Kowalski {\it et al.}\cite{Kow08}:

\begin{equation}
m_{V,rest}-M_{V,rest}=5\log _{10}[d_L(z)({\rm Mpc})]+25
.\end{equation}
In the first two static models, if we wanted to include the extinction, we should sum
$A_{V,rest}(z)=\frac{1.086c\ a_V}
{H_0(\alpha +m)}[(1+z)^m-(1+z)^{-\alpha}]$
to the distance modulus.

One aspect is remarkable: the value of the distance modulus for
the concordance model is very similar to its value for the static model
with a linear Hubble law, and it can be seen in Fig. \ref{Fig:hubble} how
the data of SNe Ia are approximately compatible with this scenario.
The fit for the concordance model over the data gives a reduced $\chi ^2$:
$\chi _r^2=3.34$;
while for the static model with linear Hubble law $\chi _r^2=4.20$, slightly
worst but not by much: the concordance model reduces the $\chi ^2$ by  only 20\%.
Lerner\cite{Ler09} also show this by comparing the residuals of the Hubble diagram in the
concordance model and in the static universe ansatz and realizing that they are
both similar. 
Is it a coincidence\footnote{The degree of coincidence depends on
the maximum redshift, $z_{max}$, we use. For instance, 
the value of $\Omega _\Lambda$ in a flat Universe which 
gives a best minimum square fit to 
the Hubble diagram for a static model with linear Hubble law and 
without extinction is $\Omega _\Lambda \approx 
0.39+0.108z_{max}-0.0085z_{max}^2$.
For $z_{max}=6$, as in our plot of Fig. \ref{Fig:hubble}, it gives
$\Omega _\Lambda=0.74$, but it falls to 0.55 if we only consider
it up to redshift 1.7, as usually for SNe Ia data. 
If we set $\Omega _m=0.3$ and we searched for the best value
of $\Omega _\Lambda $ with any value of $\Omega _{total}$, we would
get the best fit for $\Omega _\Lambda \approx 
0.76+0.171z_{max}-0.053z_{max}^2+0.0038z_{max}^3$. For $z_{max}=6$, 
it gives the best fit for $\Omega _\Lambda=0.70$, 
and it increases to 0.91 if we only consider
it up to redshift 1.7.}? With a slight extinction of 
$a_V\sim 1-2\times 10^{-4}$ Mpc$^{-3}$ (the range of the best fit
obtained in \S \ref{.extinc}), the agreement is also quite conspicuous
at least for $z<1$. For the 
simple tired light model, only with extinction of the order 
$a_V\sim 5\times 10^{-4}$ Mpc$^{-3}$ (not very far from the best fit
obtained in \S \ref{.extinc}) would get the coincidence for $z<1$.
For the plasma redshift tired light model without extinction, 
the agreement with SNe Ia for $z<1$ is also acceptable, as noted by 
Brynjolfsson\cite{Bry04a}.
This means that, with the static models,
we can fit nearly the same Hubble diagrams as
the concordance model with its cosmological constant, particularly
for supernovae fits. It is not the 
place here to extend the discussion on
the analysis of the compatibility of the static model with a linear Hubble law and
the supernovae Ia data; this would require a discussion of the systematic
errors, the selection effects, etc. At present,
I just want to emphasize that there are no major problems to make compatible
the static model of linear Hubble law with SNe Ia data.

For other cosmologies, and without extinction,
the difference from the concordance model is larger,
especially for the highest redshifts. This does not mean that they are discarded,
because the objects used as standard candles (particularly supernovae)
might have an absolute magnitude which is not constant with redshift
or some extinction along their lines of sight.
Several authors\cite{Dom00,Agu00,Goo02,Row02,Sha02,Pod06,Bal06,And06,Sch07} also
think that SNe Ia data used to derive $\Omega _\Lambda =0.7$
are affected by several systematic uncertainties that make the
$\Lambda $-CDM cosmology uncertain.

\subsection{Evolution, and other tests}
\label{.statothers}

Another topic to discuss is the apparent evolution of galaxies
at different redshift. I have not included
any evolution correction for the static models in the plots of
Fig. \ref{Fig:angsizeext} although
some slight evolution might be compatible with them, 
in the sense that brighter
populations are present at higher redshifts.
Some evolution might be present because,
as discussed in \S \ref{.results}, elliptical galaxies 
get lower angular sizes at high redshifts
than disk galaxies, possibly due to the different mean ages
of their populations or merger rate; although some of these differences 
might be due to systematic errors too, as said in \S \ref{.select}.
Also, the most massive galaxies present a higher ratio of
angular size with respect to local galaxies\cite{Tru07,van08,Bui08}, 
which may be interpreted here as
a real higher compactness with respect to the least massive galaxies.
Note, however, that, as said above, this extra-compactness can be understood in terms 
of fluctuations due to noise preventing the recovery of the extended low surface 
brightness halos in the light profile\cite{Man09}.

There is also a slight color at rest evolution\cite{Lop09} and a mass--luminosity ratio
evolution, as said in \S \ref{.brighter}; but these mass calculations are subject to important
errors depending on the synthesis model, IMF assumption, etc.; hence, there is a wide
range of possible values of mass--luminosity ratio evolution. The bluer (at rest) color of 
high redshift galaxies might be due to bias, because bluer galaxies with
younger populations are brighter. We must also bear in mind that photometric errors are
larger at higher redshift (because of the fainter fluxes), the
photometric redshifts have higher uncertainties and consequently
might affect the determination of the luminosities at rest, etc.
At present, with the data used from Refs. \cite{Tru06,Rud06} at $z\ge 1$, 
the correlation between $z$ and $(B-V)_{rest}$ is: 
$\langle z(B-V)_{rest}\rangle - \langle z\rangle \langle (B-V)_{rest}\rangle
=-0.048\pm 0.021$ for elliptical galaxies and $-0.033\pm 0.015$
for disk galaxies. The variation in color with the variation in redshift is 
much lower than the dispersion of colors and possible systematic effects.
Therefore, we cannot exclude the possibility that the luminosity evolution is small enough to be within the
error bars of our data with the static models.

Note however that the evolution of some quantities per unit comoving volume
[for instance, the star formation ratio, expressed in mass per unit
time per unit comoving volume, is claimed to be 
significantly higher in the past\cite{Hop04}] must be corrected with
respect to the concordance cosmology with Eqs. (\ref{vol1}), (\ref{vol2}), 
which reduces by a factor of 10 at $z=2$ the ratio in the static model with a linear
Hubble law with respect to the concordance one.

In any case, there may be a real evolution which should be explained, either
in expanding or static models. Note that ``static'' does not mean ``no evolution'';
``static'' means ``no expansion''; there is not necessarily a contradiction in observing
evolution in a static Universe, provided that the creation of galaxies is not
a continuous process.

Explaining the cosmic microwave background radiation (CMBR) and its anisotropies
is not the purpose of the present paper. Note, however, that there exist alternative
scenarios to explain them apart from the model of standard hot Big Bang 
(e.g., \cite{Nar03,Nar07}; see the discussion in another paper of mine\cite{Lop08}, \S 1). 
Concerning other arguments/tests in favor of the expansion (e.g.,
\cite{Lub01,Gol01,Mol02}), we must bear in mind that they
are usually a matter of discussion. Tests such as the time dilation
in SNe Ia, which were claimed to be a definitive proof of the expansion
of the Universe, find counterarguments and criticisms from opponents\cite{Bry04b,Lea06,Cra09}, 
who claim that a static Universe is compatible with the
data. The same thing happens with any other test, including
the present one of angular size. Apart from the present angular size test,
there are other tests that also present results in 
favor of a static Universe and against an expanding Universe
(e.g., \cite{LaV86,Mol91,Jaa93,Tro93,Tro96,And01,And06,Ler06,Ler09}). 

Perhaps the most immediate problem with the static Universe
is understanding the cause of the redshift of galaxies, but
there are several proposals for alternative mechanism to produce
redshifts without expansion or Doppler effect, 
so the hypothesis of a static Universe
is not an impossible one. Other facts, such as the formation of 
the large scale structure, the creation of the light elements, 
etc., also provide alternative explanations different from the standard model.
Further discussion of all the questions raised in this paragraph are given in my 
review\cite{Lop03}.

\subsubsection{UV surface brightness test}
\label{.uv}

The main discrepancy in the different tests such as angular size, surface brightness,
Hubble diagram, etc., is the evolution of galaxy luminosities, 
which is very large for the defenders of the expansion and not so 
large for the defenders of the static
models. Lerner\cite{Ler06} proposes a test of the evolution hypothesis 
that is also useful in the present case. There is a limit on the UV surface 
brightness of a galaxy, because when the surface density of hot bright stars and
thus supernovae increases large amounts of dust are produced to absorb 
all the UV except that from a thin layer.
Further increase in surface density of hot bright stars beyond a given
point just produces more dust, and a thinner surface layer, not an
increase in UV surface brightness. Based on this principle, there
should be a maximum surface brightness in UV-rest wavelengths independent
of redshift. Scarpa {\it et al.}\cite{Sca07} measured in low redshift galaxies 
a maximum FUV (1550 \AA \ at rest) emission of 18.5 mag$_{AB}$/arcsec$^2$
(the average is $24-25$ mag$_{AB}$/arcsec$^2$); no galaxy should
be brighter per unit angular area than that. 

\begin{figure*}
\vspace{1cm}
{\par\centering \resizebox*{5cm}{5cm}{\includegraphics{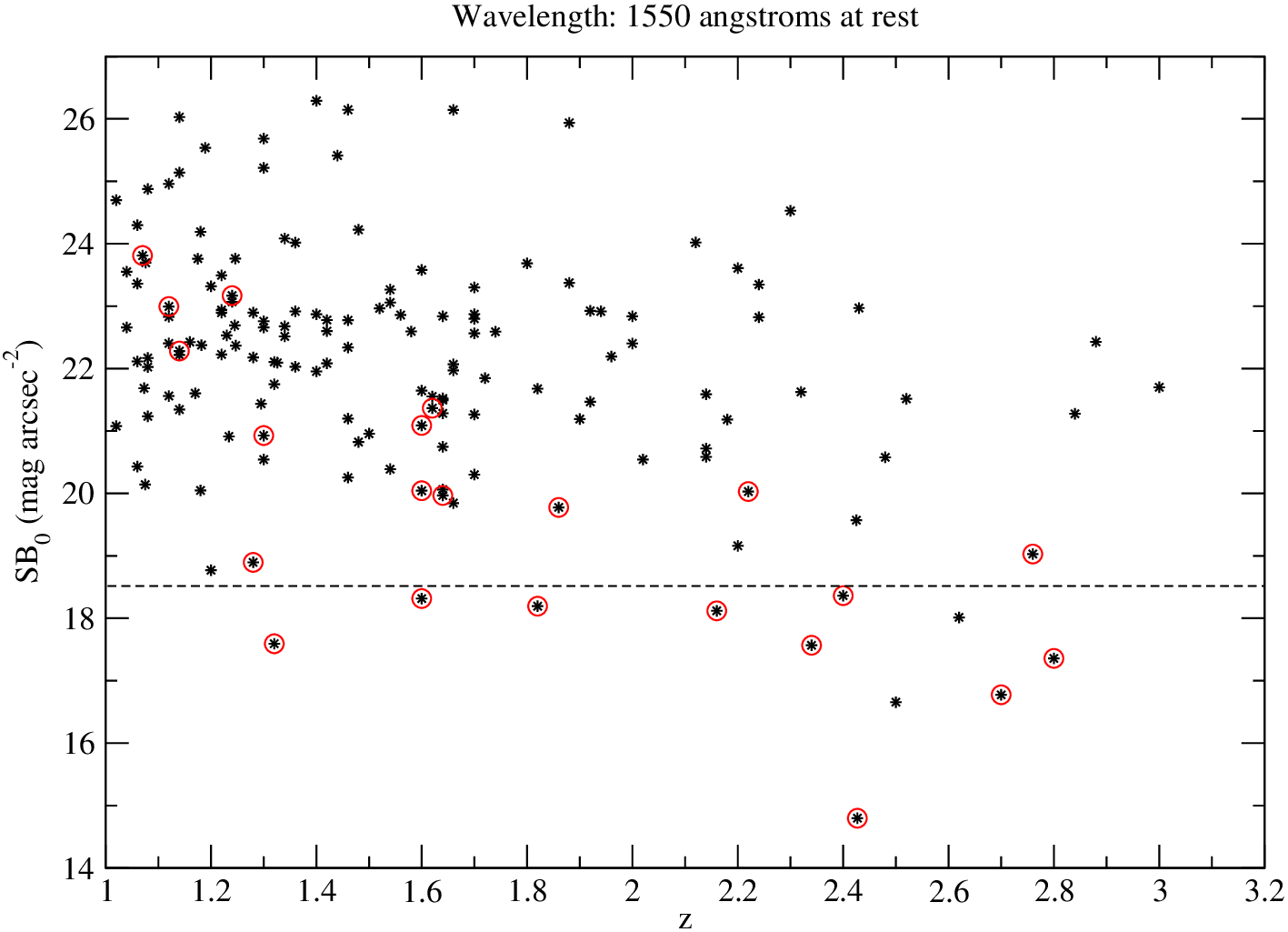}}
\hspace{1cm}\resizebox*{5cm}{5cm}{\includegraphics{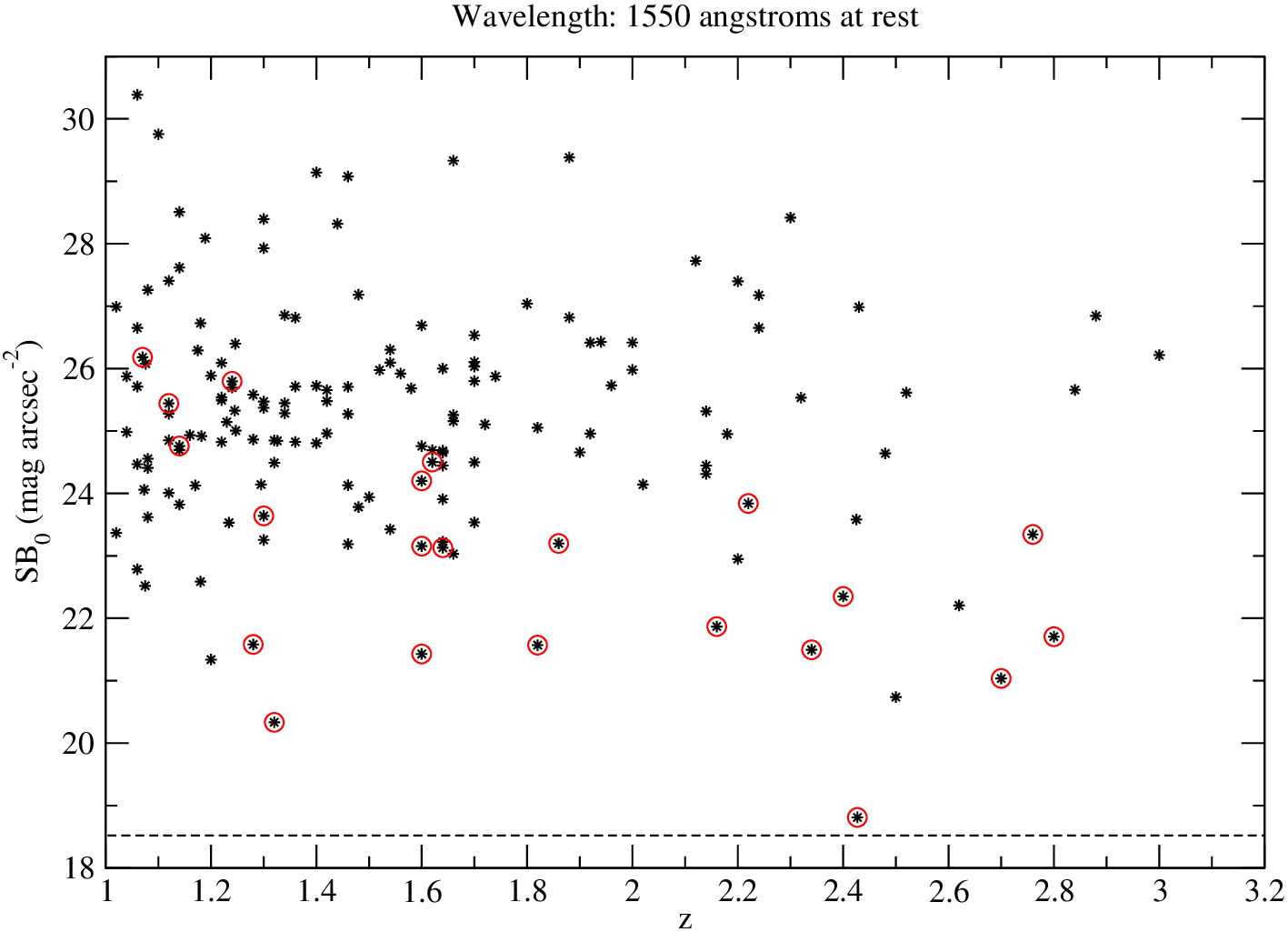}}\par}
\caption{Intrinsic FUV-rest surface brightness
of the galaxies in the subsample MS 1054-03 in an expanding (left) or static
(right) universe. 
The dashed line stands for the minimum value
of 18.5 mag/arcsec$^2$ over which all galaxies should be located.
Points with circles stand for data with $\theta <0.125"$.}
\label{Fig:surfaceb}
\end{figure*}

Using eqs. (\ref{sb}), (\ref{sb0}) with 
the flux at wavelength 1550 \AA \ at rest\footnote{An 
interpolation with the publicly available fluxes in filters U, B, V, 
I$_{814}$ and J was used to get it. These fluxes are given
in flux per unit frequency so the dimming factor must be multiplied
by a factor $(1+z)$ with respect to the flux in the whole filter.
That is, $SB_0=SB(1+z)^3$ for the expanding case, and
$SB_0=SB$ for the static case.} from the subsample
MS 1054-03 in Trujillo {\it et al.}\cite{Tru06} galaxies, and
the angular sizes $\theta _{1550 \AA }=1.14\theta _V$ \cite{McI05} 
for $n_S>2.5$ and $\theta _{1550 \AA }=1.10\theta _V$ \cite{Bar05} 
for $n_S<2.5$, I get  
the values of $SB_0$ for all the galaxies in this subsample
for the expanding or static cases (Fig. \ref{Fig:surfaceb}).
For the expanding universe, 
many galaxies have average intrinsic surface brightness ($SB_0$)
lower(brighter) than 18.5 mag$_{AB}$/arcsec$^2$, 
the galaxy MS 1054-03/1356 being the brightest one per unit angular 
area: 14.8 mag$_{AB}$/arcsec$^2$ (30 times brighter than the limit).
The angular size of this galaxy MS 1054-03/1356 (0.027$''$ circularized in V;
0.031$''$ in FUV) might be affected by some error since it is below 0.125$''$ in V, 
but even an error of 100\% in angular size would produce an error of 1.5
magnitudes in surface brightness, not 3.7 magnitudes as we observe
here. The dispersion (r.m.s.) of 
$\frac{\theta _{1550 \AA }}{\theta _V}$ 
might be around 20\% (\cite{Bar05}, Fig. 2), which means an 
uncertainty of around 0.4 mag/arcsec$^2$ (1-$\sigma $) in $SB_0$, 
much lower than the differences between $SB_0$ and its limit of 18.5.
Moreover, even avoiding galaxies with angular size less than 0.125$''$, 
there are some galaxies with surface brightness over the limit, 
up to 6 times brighter than the limit. Too high intrinsic FUV surface
brightness. Lerner\cite{Ler06} also argues why other 
alternative explanations (lower production of dust at high redshift, winds 
or others) are not consistent.
However, for the static models, all galaxies have average intrinsic 
surface brightness ($SB_0$) within $\theta $
higher(fainter) than the limit 18.5 mag$_{AB}$/arcsec$^2$, a result which
may be interpreted again in favor of the static scenario.

\subsection{Is a static model theoretically impossible?}

Apart from the discussion on the observations, which are inconclusive,
a static model is usually rejected by most cosmologists on the grounds
of a belief/prejudice that a static model is impossible.  
However, from a purely theoretical point of view, without
taking into account the astronomical observations, the representation
of the Cosmos as Euclidean and static is not excluded. Both expanding
and static space are possible for the description of the Universe, even with
evolution.

Before Einstein and the rise of Riemannian and other non-Euclidean
geometries to the stage of physics, attempts to describe the
known Universe with a Euclidean Universe were given, but with
the problem of justifying a stable equilibrium. Within a relativistic context,
Einstein\cite{Ein17} proposed a static model including a cosmological 
constant, his biggest blunder according to himself, to avoid a collapse.
This model still has problems to guarantee the stability, but it
might be solved somehow. Narlikar \& Arp\cite{Nar93} solve it
within some variation of the Hoyle--Narlikar conformal theory
of gravity, in which small perturbations of the flat Minkowski 
spacetime would lead to small oscillations about the line element 
rather than to a collapse. 
Boehmer {\it et al.}\cite{Boe07} analyze the stability of the Einstein static 
universe by considering homogeneous scalar perturbations in the context 
of $f(R)$ modified theories of gravity and it is found 
that a stable Einstein cosmos with a positive cosmological constant 
is possible. 
Other authors solve it with the
variation of fundamental constants \cite{Van84,Tro87}.
Another idea by Van Flandern\cite{Van93} is that hypothetical gravitons responsible
for the gravitational interaction
have a finite cross-sectional area, so that they can only travel a finite 
distance, however great, before colliding with another graviton. 
So the range of the force of gravity would necessarily be limited in this way
and collapse is avoided.

As said in \S \ref{.expansion}, the very concept of space expansion has its own
problems\cite{Fra07,Bar08}. The curved geometry (general relativity and its 
modifications) has no conservation of the energy-momentum of the gravity field 
(the well-known problem of the pseudo-tensor character of the energy-momentum of 
the gravity field in general relativity). However, Minkowski space follows the
 conservation of energy-momentum of the gravitational field. One approach with a 
material tensor field in Minkowski space is given in Feynman's gravitation\cite{Fey95}, where 
the space is static but matter and fields can be expanding
in a static space. It is also worth  mentioning a model related to modern 
relativistic and quantum field theories of basic fundamental interactions 
(strong, weak, electro-magnetic): the relativistic field gravity theory and fractal matter
distribution in static Minkowski space\cite{Bar08b}.

Olber's paradox for an infinite Universe also needs subtle
solutions, but extinction, absorption and reemission of light, fractal
distribution of density and
the mechanism which itself produces the redshift of the galaxies might
have something to do with its solution.
These are old questions discussed in many classical books on
cosmology (e.g., \cite{Bon61}, ch. 3) and do not warrant further 
discussion here.

\section{Conclusions}

Summing up, the main conclusions of this paper are the following:

\begin{itemize}

\item The average angular size of the galaxies for a given luminosity
with redshifts between $z=0.2$ and 3.2 is approximately
proportional to $z^{-\alpha }$, with
$\alpha $ between 0.7 and 1.2, depending on the assumed cosmology.

\item Any model of an expanding Universe without evolution is totally
unable to fit the angular size vs. $z$ dependence. 
The hypothesis that galaxies which formed earlier have much
higher densities does not work because it is not observed here that the smaller
galaxies are precisely those which formed earlier; in any case, the galaxies
observed today were formed mostly at redshifts not very different from the
galaxies observed at higher redshifts.
A very strong evolution in size would be able to get an agreement with the data but
there appear caveats to justify it in terms of age variation of the
population and/or mergers and/or ejection of massive outflows in the
quasar feedback. An average of the necessary 
two to four major mergers per galaxy during
its lifetime is excessive, and neither is it understood how
massive elliptical galaxies may present passive evolution in this scenario
or how spiral galaxies can become larger during their lifetimes.
The depletion of gas in ellipticals by a QSO feedback mechanism
does not appear to be in agreement with the observed star formation and other 
facts. Moreover, no evolution is observed in the rotation/dispersion 
velocities and, the FUV surface brightness turns out to be prohibitively high in some
galaxies at high redshift.

\item Static Euclidean models with a linear Hubble law or simple tired light
fit  the shape of the
angular size vs. $z$ dependence very well: there is a difference 
in amplitude of 20--30\%, which is within the possible systematic errors.
An extra small intergalactic extinction may also explain this difference
of 20--30\% . Some weak evolution of very high redshift sources is allowed, 
although non-evolution is a possible solution too. For the plasma redshift
tired light static model, a strong (albeit weaker than in expanding models) 
evolution in galaxy size is necessary to fit the data.
The SNe Ia Hubble diagram can also be explained in terms of
these models.

\item It is also remarkable that the explanation of test 
results with an expanding Universe requires four coincidences:

\begin{enumerate}

\item The combination of expansion and (very strong) size evolution gives 
nearly the same result as a static Euclidean universe 
with a linear Hubble law alone: $\theta \propto z^{-1}$.

\item This hypothetical evolution in size for galaxies is the same
in normal galaxies as in QSOs, as in radio galaxies, as in
first ranked cluster galaxies, as the separation among bright 
galaxies in clusters. Everything evolves in the same way to 
produce approximately a dependence $\theta \propto z^{-1}$.

\item The concordance model gives approximately the same 
(differences of less than 0.2 mag within $z<4.5$) distance modulus 
in a Hubble diagram as the static Euclidean universe with a linear Hubble law.

\item The combination of expansion, (very strong) size evolution,
and dark matter ratio variation gives the same result for the velocity
dispersion in elliptical galaxies (the result is that it is nearly constant
with $z$) as for a simple static model with no evolution in size and no 
dark matter ratio variation.

\end{enumerate}

These four coincidences might make us think that possibly we
should apply Occam's razor {\it ``Entia non sunt multiplicanda 
praeter necessitatem''}, we should use the simplest models 
that can reproduce the same things as a complex model with 
many more free parameters does.

\end{itemize}

It would be an irony of 
fate that, after all the complex solutions pursued by cosmologists 
for the last century, we had to come back to simple scenarios such as a
Euclidean static Universe without expansion. None the less, we cannot 
at present defend any of these simple models apart from the standard one
because this would require other analyses. The conclusion of this paper
is just that the data on angular size vs. redshift 
present some conflict with the standard model, and that they are
in accordance with a very simple phenomenological extrapolation of the Hubble
relation that might ultimately be linked to a static model of the universe.
 
\section*{Acknowledgments}
Thanks are given to: the anonymous referee for very helpful suggestion to
improve this paper;
Ignacio Trujillo and Eric J. Lerner
for useful discussions on many topics related to this paper
and comments on a draft of it; Riccardo Scarpa, 
Carlos M. Guti\'errez, Juan Betancort-Rijo also 
for their comments on a draft of this paper; Daniel
MacIntosh for providing me the GEMS data from the
papers \cite{McI05,Bar05}; Terry J. Mahoney for proof-reading this paper.
MLC was supported by the {\it Ram\'on y Cajal} Programme
of the Spanish Ministry of Science.

\end{document}